%% file: main.tex
\documentclass[letterpaper,twocolumn,10pt]{article}
\usepackage{hyperref} 
\usepackage{usenix}
\ifXeTeX\microtypesetup{spacing=false,kerning=false}\fi
\usepackage{amsmath,amssymb,graphicx,booktabs,array,tabularx,float,caption}
\usepackage{placeins}
\usepackage{afterpage}
\usepackage{enumitem}
\usepackage{balance}
\hypersetup{hidelinks,pdftitle={DHSched: Stateless Control for Stateful Real-Time Avatar Serving},pdfauthor={}}
\floatstyle{plaintop}\newfloat{algorithm}{tbp}{loa}\floatname{algorithm}{Algorithm}
\setlist[itemize]{leftmargin=*,itemsep=2pt,topsep=4pt,parsep=0pt}
\makeatletter
\def\@maketitle{\newpage
  \begin{center}%
    {\LARGE\bfseries\@title\par}%
    \vskip 1.4em
    {\large\@author\par}%
  \end{center}%
  \vskip 0.9em}
\makeatother
\title{DHSched: Stateless Control for Stateful Real-Time Avatar Serving}
\author{%
\begin{tabular}{@{}c@{}}
Xin Wang\textsuperscript{1,$*$}\quad Haitong Zhang\textsuperscript{2,$*$}\quad
Xianghong Li\textsuperscript{1}\quad Shumin Lin\textsuperscript{1}\\[0.55em]
Xianzheng Song\textsuperscript{1}\quad Lin Wang\textsuperscript{1,$\dagger$}\quad
Zhenyu Xu\textsuperscript{2,$\dagger$}\\[1.05em]
{\normalsize\textsuperscript{1}AiShiWeiLai Co., Ltd., Beijing 100094, China}\\[0.15em]
{\normalsize\textsuperscript{2}College of Computer Science, Sichuan University, Chengdu 610065, China}\\[0.95em]
{\small\ttfamily wangxin03@yuaiweiwu.com,\quad 1325586035@qq.com,\quad xianghongleiking@gmail.com,}\\[0.3em]
{\small\ttfamily linshumin@yuaiweiwu.com,\quad songxianzheng@yuaiweiwu.com,}\\[0.3em]
{\small\ttfamily wanglin@yuaiweiwu.com,\quad sanxu@scu.edu.cn}\\[0.95em]
{\footnotesize\textsuperscript{$*$}Equal contribution.\quad\textsuperscript{$\dagger$}Corresponding authors.}
\end{tabular}}
\date{}
\begin{document}
\raggedbottom
\maketitle
\input{sections/00-abstract}
\input{sections/01-introduction}
\input{sections/09-related-work}
\input{sections/02-background-and-motivation}
\input{sections/03-dhsched-overview}
\input{sections/04-cross-plane-ownership-architecture}
\input{sections/05-capacity-safe-placement-of-new-generations}

\input{sections/06-source-independent-ownership-transfer}
\input{sections/07-implementation}
\input{sections/08-evaluation}
\input{sections/10-discussion}
\input{sections/11-conclusion}
\bibliographystyle{plain}
\bibliography{refs}

\end{document}

%% file: sections/00-abstract.tex
\begin{abstract}

Real-time avatar services run long-lived, GPU-backed sessions that transform a continuous text stream into speech, facial motion, rendered video, and RTC media. Elastic operation repeatedly reassigns these sessions across Workers. A reassignment updates shared placement state while the previous GPU runtime and RTC connection remain alive, opening a gap between control-plane assignment and execution authority. A stateful controller serializes ownership by retaining per-session authority, placing that state on the controller\textquotesingle s scaling and failover path.

We present \textbf{DHSched}, a control plane that manages long-lived stateful sessions through stateless peer Dispatch replicas. DHSched externalizes session ownership as a generation identified by (WorkerID, epoch). Dispatch conditionally commits each generation. Infer-Controller claims and revalidates the same generation before releasing state-advancing input. A stale runtime loses execution authority even if it remains alive. DHSched ranks candidates from periodically refreshed load views while keeping capacity admission authoritative. Recovery advances the same generation path via source-epoch guards and source-independent reconstruction.

DHSched runs in our production avatar-serving stack, which reached \textbf{49,987 concurrent live sessions}. On 10 June 2026, it completed \textbf{9,927 session migrations} with no observed dual-owner conflicts or capacity overshoots. At the production peak, CreateLive latency was \textbf{98.3 ms at P99} and PushText latency was \textbf{16.5 ms at P99}. Across \textbf{100,000 randomized ownership runs}, assignment-time, claim-only, and WorkerID-only protocols permit stale execution, while DHSched records no dual-owner or stale-generation release.

\end{abstract}

%% file: sections/01-introduction.tex
\section{Introduction}\label{introduction}

Digital-avatar serving couples a continuous LLM-driven input stream with GPU inference, rendering, and RTC delivery. Once a live starts, its Worker preserves inference context, rendering state, the RTC connection, and session-specific runtime across many inputs. A Worker can host hundreds of concurrent lives, making each live a long-lived, stateful, GPU-backed execution unit for fleet scheduling.

Active sessions undergo routine reassignment in response to load skew, Worker failures, scaling, preemption, and planned drain. Concentrated load increases GPU queueing and delays media generation, making reassignment a normal part of maintaining service quality and session availability.

Reassigning a stateful session shifts execution responsibility along with placement. When Dispatch redirects a live from Worker A to Worker B, shared state may already name B while A still retains its runtime and RTC connection. If both Workers continue to receive input, they advance the same live and emit conflicting output. During handoff, a brief zero-owner interval is acceptable, provided that at most one Worker remains authorized to consume new input. Assignment and execution authority change at different points in the handoff.

At production scale, these handoffs are coordinated by a horizontally scaled control plane. Dispatch runs as stateless peer replicas, and operations on the same live, including creation, retries, reassignment, drain, and recovery, may be handled by different replicas over time. A Dispatch request completes after its control update, while Worker execution and input delivery continue across subsequent operations. Keeping per-live authority inside Dispatch makes controller-local state part of the session lifetime and its failover path.

\textbf{How can stateless peer controllers safely manage long-lived, stateful real-time sessions?}

Request-level LLM serving systems such as vLLM~\cite{kwon2023vllm},
Orca~\cite{yu2022orca}, AlpaServe~\cite{li2023alpaserve}, and
Sarathi-Serve~\cite{agrawal2024sarathi} optimize model execution through
batching, KV-cache management, and GPU scheduling. Their primary scheduling
unit is an inference request or model execution context within the serving
runtime.

GPU migration systems extend scheduling beyond initial placement.
Llumnix~\cite{sun2024llumnix} migrates KV-cache state for in-flight requests,
while TurboServe~\cite{jiang2026turboserve} combines placement, rebalancing,
and runtime-state migration for long-lived streaming workloads. Their transfer
paths move the execution state needed to continue the workload at the destination.

Stateful placement systems address ownership more directly.
Slicer~\cite{adya2016slicer} separates assignment from request forwarding and
orders ownership changes through generations and conditional updates.
Orleans~\cite{bykov2011orleans} associates stable logical identities with
movable activations through its directory and runtime coordination.

RTC-backed avatar serving places these functions on separate paths. Dispatch selects the execution target, Infer-Controller mediates input delivery, and the Worker hosts the GPU and RTC runtime. A control update can complete while the input path and an earlier Worker runtime remain active. Target selection reads periodically refreshed fleet state, and recovery can begin after the serving Worker has failed or entered drain.

DHSched carries each ownership change across these paths as a generation identified by (WorkerID, epoch). Dispatch conditionally commits the next selected Worker in shared state, while activation occurs later when the target reaches Infer-Controller. Infer-Controller checks the same generation before every state-advancing input release. An earlier runtime may survive the handoff after its generation has lost authority to advance the live. The epoch distinguishes successive assignments to the same Worker, including an A\(\rightarrow\)B\(\rightarrow\)A sequence.

Each new generation also needs a feasible target. Dispatch ranks eligible Workers from periodically refreshed Pool state. Between refreshes, concurrent replicas can make correlated choices from the same view. DHSched incorporates recent assignments observed by each replica, distributes selection across lightly loaded Workers, and preserves configured Pool priority. Cached state orders candidates, while authoritative Worker membership decides whether an assignment is admitted before the ownership transition commits.

Recovery reuses the generation transition when the serving Worker is unavailable or preparing to leave. Each recovery record carries the source epoch observed when the work was created and remains valid merely while the live is still at that generation. A newer ownership transition makes delayed or duplicate recovery work stale. For current work, DHSched selects and admits a replacement, advances the live to a higher generation, and reconstructs the target from shared session metadata and pending input without copying GPU or RTC runtime state from the source. Planned drain uses the same transition after preparing replacement capacity and reaching a suitable stream boundary.

DHSched runs in our production avatar-serving stack, which reached \textbf{49,987 concurrent live sessions} and completed \textbf{9,927 session migrations} on 10 June 2026 with no observed dual-owner conflicts or capacity overshoots. Controlled experiments isolate the role of the generation protocol. Across 100,000 randomized runs per protocol, assignment-time and claim-only variants violate single-owner execution in more than 99\% of runs. Per-release WorkerID-only validation reduces these failures but still violates single-owner execution in \textbf{25.5\% of A\(\rightarrow\)B\(\rightarrow\)A runs}. DHSched records no dual-owner or stale-generation release.

This paper makes three contributions.

\begin{itemize}
\item
  \textbf{A cross-plane single-owner protocol for stateless peer controllers.} DHSched represents live-session authority with a monotonically advancing generation identified by (WorkerID, epoch). Entity-level short locks and CAS serialize ownership changes, and epoch fencing rejects stale execution at the data plane. Each live retains at most one authorized input consumer across concurrent create, retry, reassignment, and recovery.
\item
  \textbf{Hierarchy-aware, capacity-safe placement under stale load views.} DHSched leverages bottom-tier randomization, replica-local load deltas, and cross-Pool waterlines to spread concurrent choices across eligible capacity. Final capacity admission is atomic, so stale ranking affects placement quality without authorizing oversubscription.
\item
  \textbf{Source-independent ownership transfer and recovery.} DHSched guards recovery with the source epoch, selects and admits a replacement, and commits a higher generation. The target reconstructs execution from shared state without source-side GPU or RTC runtime transfer. The same recovery design covers Worker failure, planned drain, and proactive reassignment.
\end{itemize}

%% file: sections/09-related-work.tex
\section{Related Work}\label{related-work}

\subsection{Stateful Ownership and Decentralized Placement}\label{stateful-ownership-and-decentralized-placement}

Distributed systems have long combined fine-grained placement with explicit ownership management. Centrifuge~\cite{adya2010centrifuge} couples partitioning with lease management
for in-memory state. Slicer~\cite{adya2016slicer} separates assignment from
request forwarding and orders assignment changes through monotonically increasing
generations and conditional updates; its strong-consistency path also coordinates
handoff through leases. Orleans~\cite{bykov2011orleans} associates stable logical
identities with movable activations through its grain directory and runtime
coordination.

DHSched places selection, persistent input delivery, and GPU/RTC execution in separate components, and fences each ownership generation on the input path after control-plane assignment.  The protocol can advance ownership without waiting for the previous runtime to terminate.

Epoch-based fencing also appears in Kafka consumer groups~\cite{kreps2011kafka},
where membership epochs reject stale consumers after ownership changes.
Azure Service Fabric~\cite{kakivaya2018servicefabric} maintains replicated
application state and promotes a surviving replica after failure. DHSched assumes that recoverable session metadata and pending input already reside outside the execution Worker, allowing a replacement runtime to be reconstructed after the previous Worker becomes unavailable.

Decentralized schedulers address placement with incomplete or changing load information. Power-of-two choices~\cite{mitzenmacher2001power} and systems such as
Sparrow~\cite{ousterhout2013sparrow} reduce dependence on a global load view,
while Omega~\cite{schwarzkopf2013omega} allows concurrent schedulers to make
optimistic decisions over shared state.Orleans\textquotesingle{} ActivationCountBasedPlacement combines randomized selection with cached activation counts and a local estimate of recent placements. DHSched uses a similar stale-view setting but also accounts for Pool eligibility and deployment priority, with authoritative capacity admission applied after candidate ranking.

\subsection{GPU Serving and Stateful Migration}\label{gpu-serving-and-stateful-migration}

GPU-serving systems such as Orca, vLLM, AlpaServe, and Sarathi-Serve improve model execution through batching, model placement, KV-cache management, and prefill/decode scheduling. Their scheduling unit is primarily an inference request or model execution context. DHSched schedules a longer-lived session that persists across many inference operations together with application and RTC state.

Llumnix migrates execution state for in-flight LLM requests, while TurboServe combines placement, suspension, and GPU-to-GPU migration for long-lived streaming workloads. These designs retain active work by transferring runtime state from the source. DHSched reconstructs execution from externalized session metadata and pending input, allowing ownership transfer to proceed when source participation is unavailable.

Other systems also treat long-running stateful computation as a scheduling unit. SAGA~\cite{guo2026saga} schedules agent workflows while exploiting KV reuse
and session affinity. Stateful stream-processing systems such as
Falcon~\cite{mishra2024falcon},
PAM~\cite{mishra2024pam}, and
DRRS~\cite{qing2025drrs}
coordinate reconfiguration by moving operator or key state. DHSched targets sessions whose recoverable state is already externalized from the execution Worker.

\subsection{Real-Time Avatar Serving}\label{real-time-avatar-serving}

Real-time avatar and streaming-generation systems such as
SyncAnimation~\cite{liu2025syncanimation},
Livatar-1~\cite{liu2025livatar},
Live Avatar~\cite{huang2025liveavatar},
Hallo-Live~\cite{li2026hallolive}, and
StreamDiffusionV2~\cite{feng2026streamdiffusionv2}
improve interactive generation through model architecture, streaming inference,
temporal stabilization, batching, and GPU execution optimizations. These systems primarily optimize the execution of an already placed session.

DHSched addresses the surrounding fleet-level control problem, encompassing placement, ownership transfer during failure or drain, and input fencing while runtimes from successive assignments overlap.

%% file: sections/02-background-and-motivation.tex
\section{Background and Motivation}\label{background-and-motivation}

Real-time avatar serving separates control, persistent input delivery, and GPU execution across independently scaled components. A session can pass through many control operations while its input stream and Worker runtime remain active.

\subsection{Disaggregated Avatar Serving}\label{disaggregated-avatar-serving}

A live session spans three serving paths. Dispatch handles creation, Worker selection, reassignment, and recovery. Infer-Controller carries the continuous text stream between the application and the serving Worker. GPU Workers perform avatar inference and rendering, maintain RTC connections, and hold session-local runtime state. A separate administrative path manages Worker registration, maintenance, and drain.

Dispatch operations complete within one control update, while PushText, PullText, Worker runtimes, and RTC connections persist across many updates. Shared assignment can change before the previous runtime has torn down. 

\subsection{Session State and the Ownership Boundary}\label{session-state-and-the-ownership-boundary}

The control plane stores the current assignment together with the membership, load, and capacity information used for placement. Placement state is refreshed periodically and can lag changes in Worker load and membership.

The input path retains pending text, request identifiers, acknowledgements, and replay position across creation, reassignment, and recovery. Infer-Controller continues serving this stream as the assigned Worker changes.

Execution state remains local to the Worker. It comprises the avatar runtime, GPU inference and rendering context, and RTC connection. A Worker from an earlier assignment can retain this state after a newer assignment has committed.

\subsection{Production Pressures}\label{production-pressures}

\textbf{Concurrent ownership changes.} Different Dispatch replicas handle creation, retries, reassignment, drain, and recovery for the same live. Delayed work arrives after another operation has already changed the assignment, while a runtime associated with the previous assignment keeps pulling input.

\textbf{Stale-view placement.} Dispatch replicas rank Workers from periodically refreshed Pool state. Requests arriving between refreshes can observe the same low-load candidates, causing strict least-load selection to concentrate concurrent choices around a cached minimum. Candidate ranking reflects an older fleet view than the capacity available when placement is attempted.

\textbf{Failure fan-out.} A single Worker can host many active lives. Its failure generates concurrent recovery work for those sessions, including retries or duplicate records for lives whose ownership may already have advanced. Replacement selection and runtime reconstruction proceed while the source Worker may remain unavailable.

%% file: sections/03-dhsched-overview.tex
\section{DHSched Overview}\label{dhsched-overview}

\begin{figure*}[!t]
\centering
\includegraphics[width=0.86\linewidth]{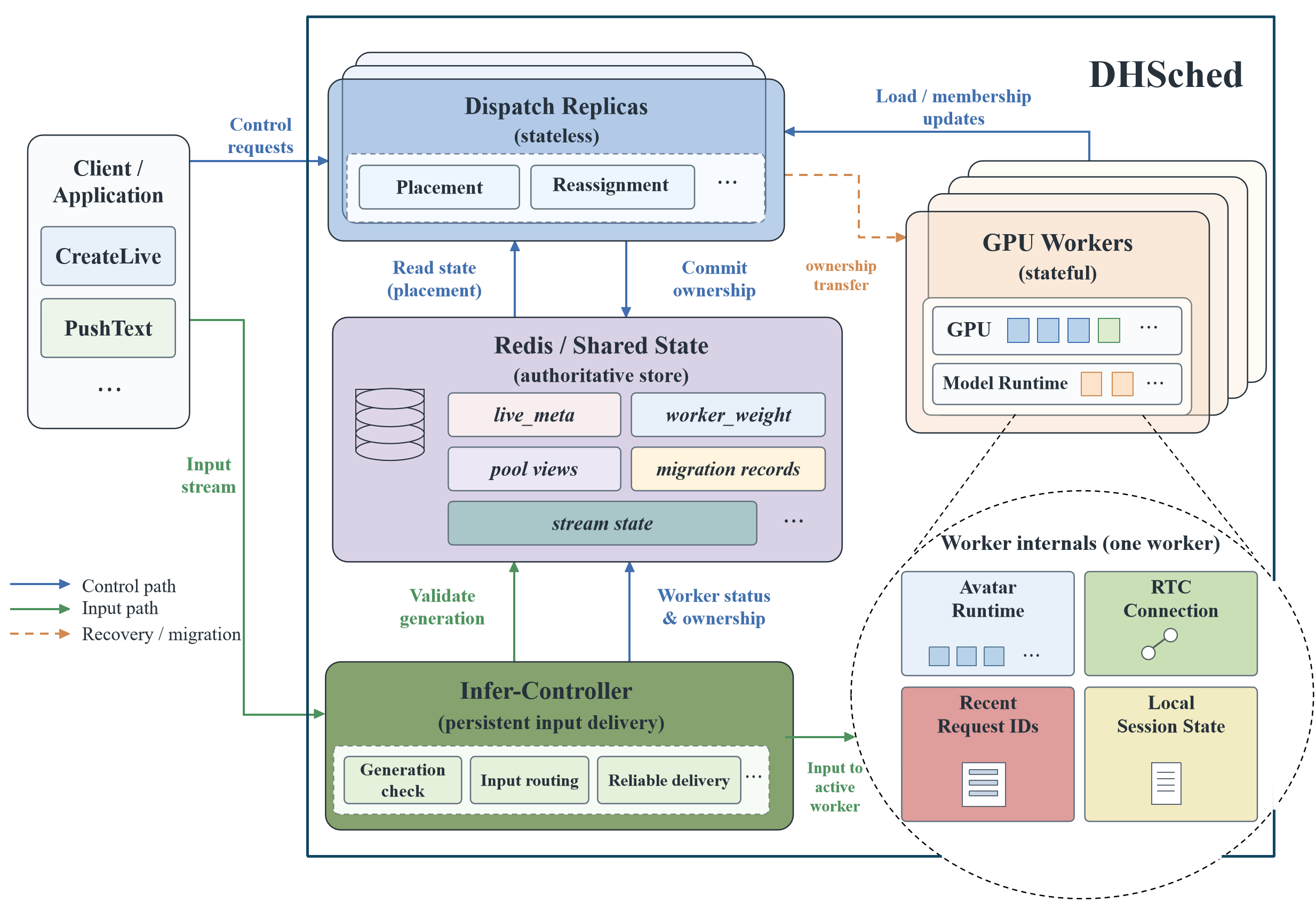}
\caption{DHSched architecture. Stateless Dispatch replicas commit shared ownership; Infer-Controller fences input to stateful GPU Workers.}
\label{fig:1}
\end{figure*}

Figure~\ref{fig:1} shows the DHSched control path and its shared-state boundary. Dispatch runs as stateless peer replicas with no elected leader or replica-to-replica ownership protocol. Authoritative ownership and capacity state are stored in Redis, while Infer-Controller and Workers retain the long-lived input and execution state described in Section~\ref{background-and-motivation}.

A session transition follows one path through these components. Dispatch selects an eligible Worker and obtains capacity admission before conditionally committing a new ownership generation (WorkerID, epoch). The selected Worker then creates or reconstructs its runtime and reaches Infer-Controller with the committed generation. Its first pull activates that generation, and later state-advancing input releases are checked against the same Worker and epoch. CreateLive establishes the first generation, while reassignment, recovery, and planned drain advance this path from an existing generation.

\textbf{Assumptions.} DHSched assumes a linearizable authoritative store for ownership and capacity transitions. Every state-advancing input release passes through Infer-Controller and is checked against the current generation. Control requests, messages, and recovery work may be delayed, duplicated, or retried.

%% file: sections/04-cross-plane-ownership-architecture.tex
\section{Cross-Plane Ownership Architecture}\label{cross-plane-ownership-architecture}

DHSched separates a control-plane assignment from the authority to consume session input. Dispatch commits the next owner in shared state. The selected Worker later claims that generation through Infer-Controller. Every subsequent state-advancing input release is checked against the same generation.

\subsection{Authority Generations and the Release Rule}\label{authority-generations-and-the-release-rule}

\begin{figure*}[tbp]
\centering
\includegraphics[width=0.78\linewidth]{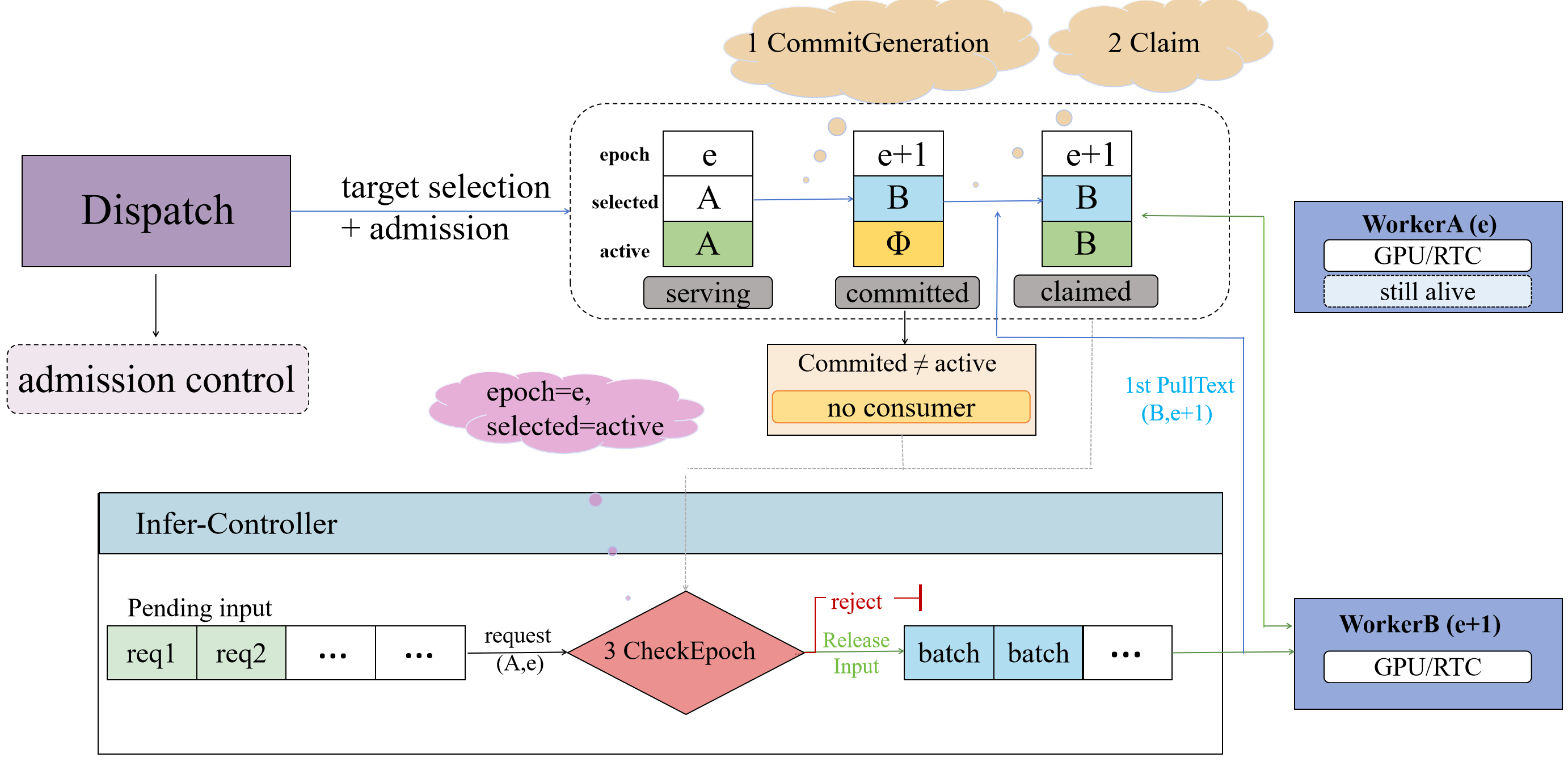}
\caption{Generation handoff across control and input paths. Commit selects the next Worker; Claim activates it; per-input checks fence the previous generation.}\label{fig:2}
\end{figure*}

Let \(S\) be the set of live sessions and \(W\) the set of Workers. For each live \(s\), the authoritative ownership record has the logical form

\begin{equation}\label{eq:1}
O(s)=\langle\mathsf{epoch}(s),\mathsf{selected}(s),\mathsf{active}(s)\rangle.
\end{equation}

selected records the Worker chosen by the control plane. active records the Worker that has completed the data-plane claim. These fields mark different points in a handoff (Figure~\ref{fig:2}). Dispatch can commit a new assignment before the target finishes runtime creation and RTC establishment, while active remains empty until the target reaches the persistent input path.

An ownership generation is identified by

\begin{equation}\label{eq:2}
g = \left\langle w,e \right\rangle,
\end{equation}

where \(w\) is the selected Worker and \(e\) is the epoch. The epoch increases whenever ownership advances to a new generation.

\begin{figure}[tbp]
\centering
\includegraphics[width=\linewidth]{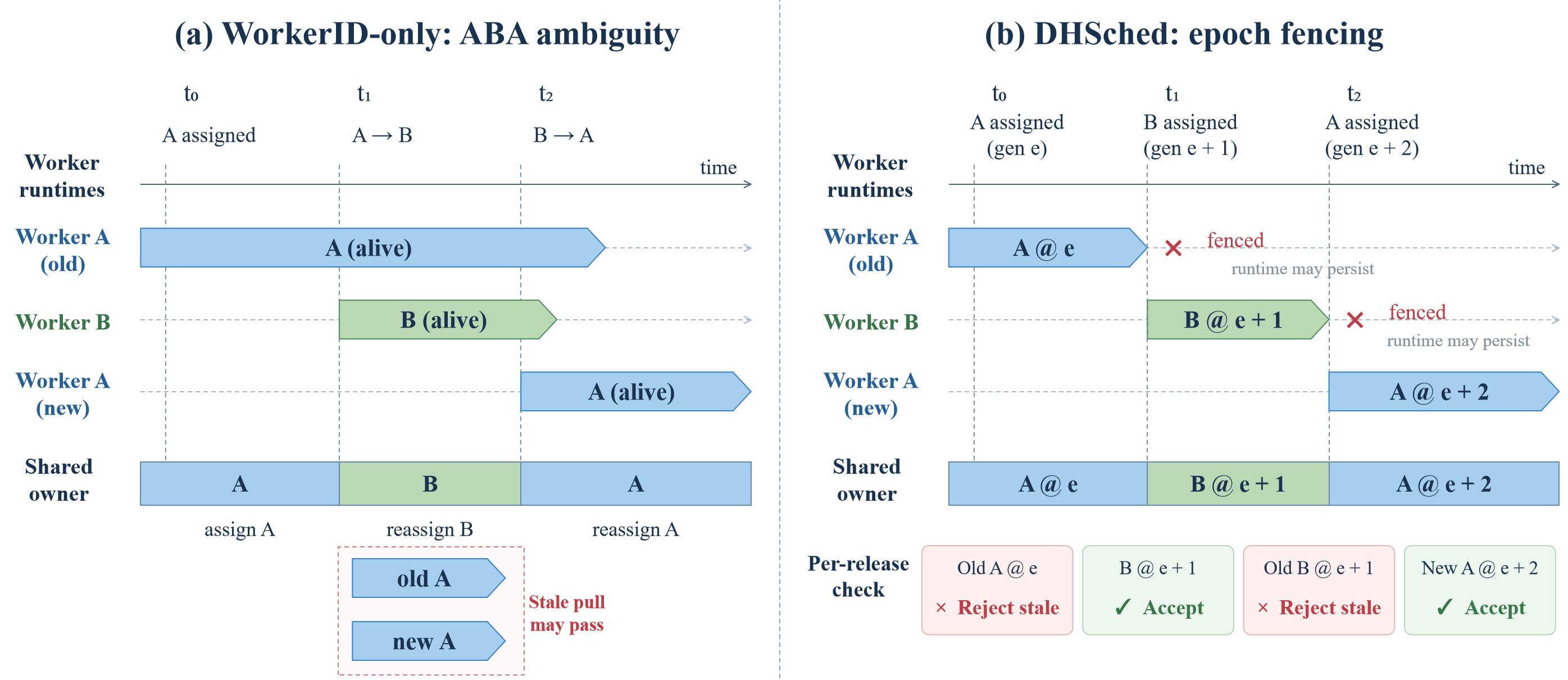}
\caption{ABA ambiguity with WorkerID-only validation and its prevention by epoch fencing. An old runtime can remain alive after losing input authority.}\label{fig:3}
\end{figure}

Worker identity alone does not identify authority. A live can move from Worker A to Worker B and later return to A. The old A runtime and the new A assignment share the same WorkerID, while belonging to different ownership generations. The epoch distinguishes these two executions (Figure~\ref{fig:3}).

Infer-Controller enforces the current generation at ReleaseInput. For a request from Worker \(w\) carrying epoch \(e\), \(\mathsf{ReleaseInput}\left( s,w,e \right)\) is enabled only when

\begin{equation}\label{eq:3}
\begin{aligned}\mathsf{epoch}(s)&=e\;\land\;\mathsf{selected}(s)=w\\&\quad\land\;\mathsf{active}(s)=w.\end{aligned}
\end{equation}

A handoff may temporarily leave active empty. Runtime existence, GPU activity, and an open RTC connection do not satisfy the release condition.

\subsection{Committing a Generation}\label{committing-a-generation}

CreateLive, reassignment, and recovery begin from the generation observed by the caller. After target selection and capacity admission, Dispatch attempts the transition in Algorithm~\ref{alg:1}.

\begin{algorithm}[tbp]
\centering
\refstepcounter{algorithm}\label{alg:1}
\includegraphics[width=\linewidth]{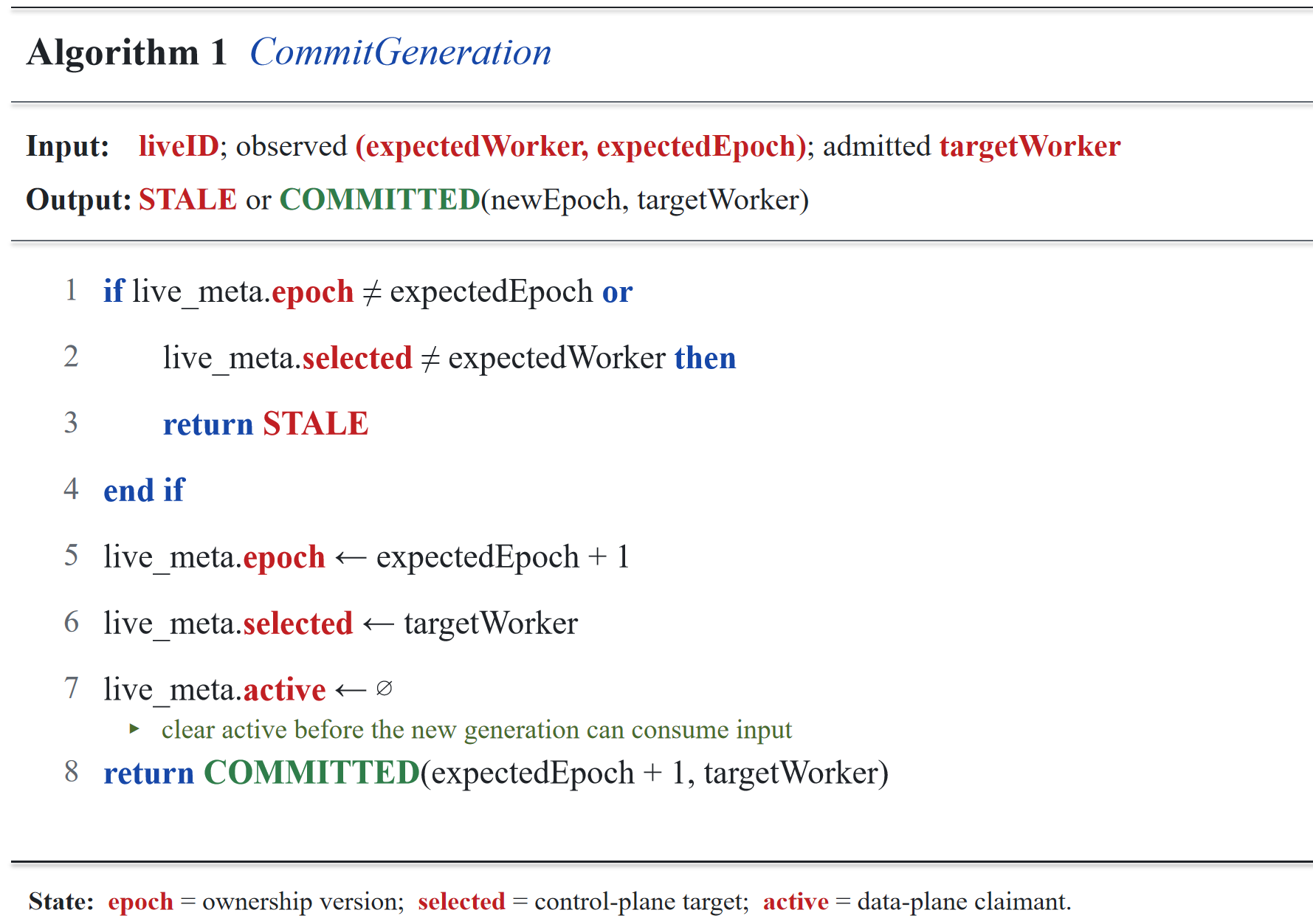}
\end{algorithm}

The conditional update is the ordering point for ownership. Once one operation advances epoch \(e\), another operation that still expects \(e\) returns STALE without modifying the new generation.

The same update clears active. The commit records the next control-plane owner without immediately granting input authority. The target acquires that authority only after reaching Infer-Controller and completing the claim in Section~\ref{claim-and-per-item-fencing}.

Dispatch uses a short per-live lock around the surrounding multi-step work, including target selection and retry handling. The lock reduces redundant concurrent work. The conditional generation update orders ownership. Data-path checks enforce input authority after the lock is released.

The commit also updates epoch-tagged desired state and the target\textquotesingle s capacity membership. These records drive runtime activation and later cleanup. They carry no input authority.

A Dispatch replica can crash after the commit and before returning its response. A retry reads the committed generation and returns the existing result. An operation superseded by a newer generation returns STALE.

\subsection{Claim and Per-Item Fencing}\label{claim-and-per-item-fencing}

After the generation commit, the selected Worker creates or reconstructs its local runtime and joins the RTC room. Its first PullText carries (liveID, WorkerID, epoch). Infer-Controller executes the claim atomically (Algorithm~\ref{alg:2}).

\begin{algorithm}[tbp]
\centering
\refstepcounter{algorithm}\label{alg:2}
\includegraphics[width=\linewidth]{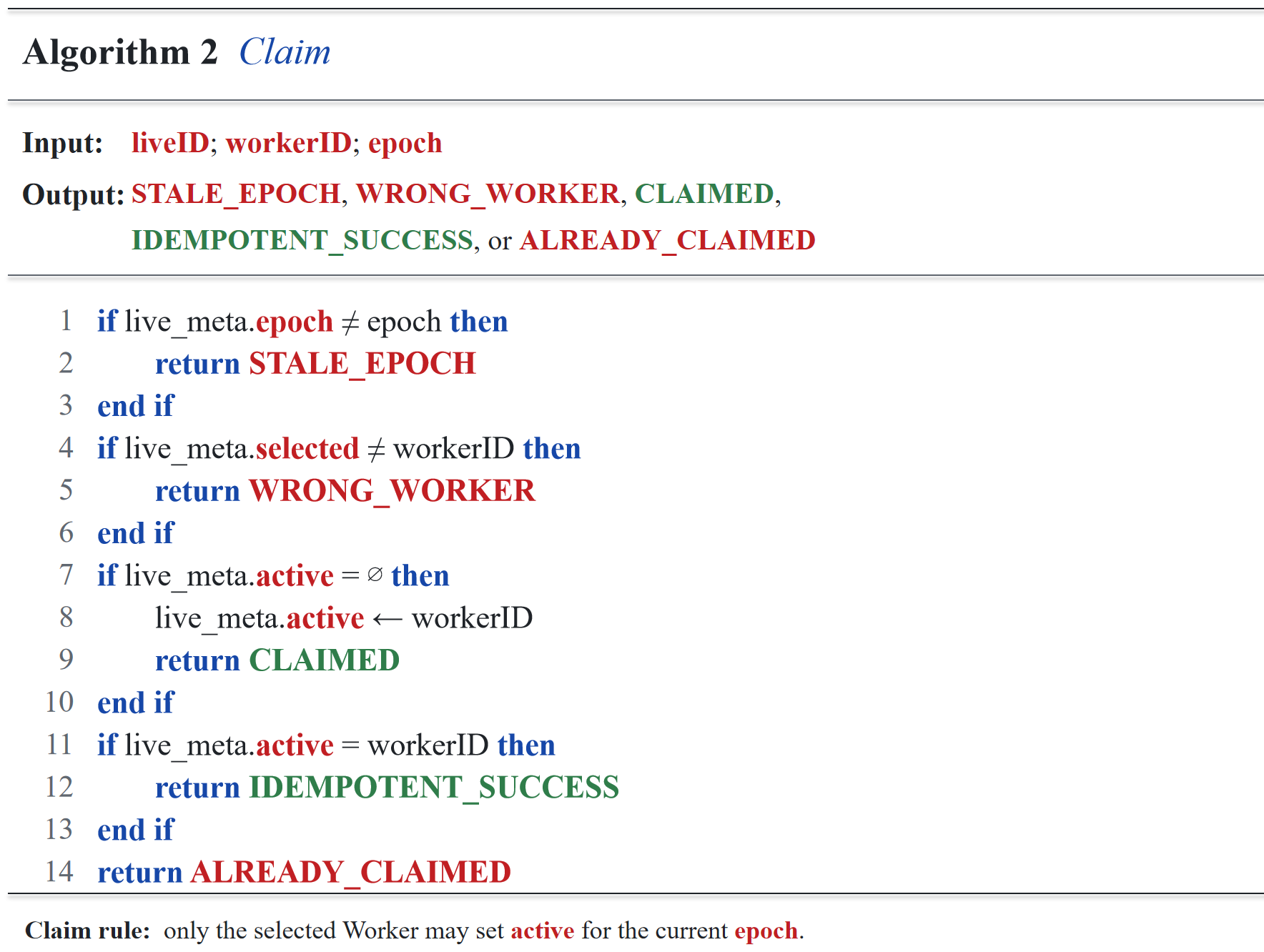}
\end{algorithm}

The first successful claim fixes the active Worker for that generation. Claims from another Worker fail. Repeated claims from the selected Worker leave the ownership state unchanged.

A claim is valid only for its generation. The connection can survive a later reassignment, so Infer-Controller revalidates the generation before each state-advancing release.

Once a higher generation commits, subsequent releases from the previous generation fail this check. The old GPU runtime or RTC connection can remain alive until local teardown completes. It receives no new state-advancing input.

This check covers every path that advances the live, including PullText and equivalent downlink operations that release new text or control commands.

The persistent input path is the fencing boundary. Dispatch does not wait for the previous Worker to acknowledge teardown before committing the next generation.

\subsection{Single-Owner Safety}\label{single-owner-safety}

Under the storage and data-path assumptions in Section~\ref{dhsched-overview}, DHSched enforces

\begin{equation}\label{eq:4}
\begin{gathered}\forall s\in S,\\\left|\left\{(w,e)\mid\mathsf{ReleaseInput}(s,w,e)\text{ is enabled}\right\}\right|\leq1.\end{gathered}
\end{equation}

The invariant follows from the three operations that establish or exercise authority.

CommitGeneration. A successful commit advances the epoch, installs one selected Worker, and clears active in one ordered transition. The new generation begins with no enabled ReleaseInput.

Claim. Claim changes active only from empty to the Worker already selected in the same epoch. Atomic execution allows one such value for the generation.

ReleaseInput. Every release checks the current epoch, selected Worker, and active Worker. Consider a release from generation \(e\) racing with a commit of \(e + 1\). A release ordered before the commit belongs to generation \(e\). After the commit, the changed epoch and cleared active reject every later release carrying \(e\). The replacement cannot release input until it claims \(e + 1\).

The initial state has no active Worker. Each transition preserves at most one enabled release tuple for a live, so the property holds over every reachable transition sequence.

The guarantee applies to authority over future input. Input transport retains the service\textquotesingle s existing at-least-once semantics. A request released before a generation change can still be retried after a lost acknowledgement, with bounded request-ID deduplication at the Worker.

\subsection{Epoch-Guarded Idempotence and No Resurrection}\label{epoch-guarded-idempotence-and-no-resurrection}

Every operation that can modify ownership carries the generation on which it was based. Create and reassignment include the expected epoch. Recovery records carry (liveID, sourceEpoch). Cleanup state retains the epoch or reservation identifier that created it.

DHSched applies four rules.

\begin{itemize}
\item
  A retry of a completed operation returns the generation already committed.
\item
  An operation can advance ownership only from the generation it observed.
\item
  Work carrying an older epoch performs no ownership update.
\item
  Cleanup removes only state tagged with the matching epoch or reservation identifier.
\end{itemize}

A stale recovery item is discarded after another transition advances the epoch, and cleanup from an earlier generation cannot remove state tagged by a newer one.

The epoch is required even when the selected Worker is unchanged across nonadjacent generations. In an A\(\rightarrow\)B\(\rightarrow\)A sequence, work created by the first A assignment remains stale after ownership returns to A because it carries the earlier epoch.

\subsection{Formal Model}\label{formal-model}

Our TLA+ specification~\cite{lamport2002specifying} models epoch, selected,
and active directly. CommitGeneration, Claim, and ReleaseInput follow the transitions above. Each successful ReleaseInput(s,w,e,item) is added to an observable trace only when the full release guard holds.

Extension variables model capacity reservations, desired Worker membership, and recovery work. Their actions include Reserve, EnqueueRecovery, Recover, CancelReservation, and Reconcile.

TLC checks five invariants.

\noindent\textbf{I0 --- MonotonicEpoch}

epoch{[}s{]} never decreases.

\noindent\textbf{I1 --- UniqueGeneration}

A live has at most one selected Worker in a given epoch.

\noindent\textbf{I2 --- SingleOwner}

At most one (Worker, epoch) pair per live can satisfy the input-release guard in any reachable state.

\noindent\textbf{I3 --- NoResurrection}

An action carrying epoch \(e\) cannot modify ownership after the live has advanced beyond \(e\).

\noindent\textbf{I4 --- CapacitySafety}

For every Worker \(w\),

\begin{equation}\label{eq:5}
\mathsf{committed}(w) + \mathsf{reserved}(w) \leq \mathsf{capacity}(w).
\end{equation} TLC explores races among generation commits, claims, input releases, stale recovery work, reservation cleanup, and reconciliation. SingleOwner is checked over both reachable release guards and the observable release trace.

The liveness model assumes weak fairness for retries and the continued availability of at least one eligible Worker with free capacity. Under these assumptions, an unowned live eventually reaches a committed generation, abandoned reservations are eventually reclaimed, and current recovery work eventually leaves the queue.

%% file: sections/05-capacity-safe-placement-of-new-generations.tex
\section{Capacity-Safe Placement of New Generations}\label{capacity-safe-placement-of-new-generations}

Every new generation needs an eligible Worker with available capacity. Dispatch makes this choice from a cached view of Pool membership and Worker load. The view is refreshed periodically and lags recent assignments during a burst. DHSched keeps candidate selection local to each Dispatch replica and checks capacity against authoritative shared state before committing the generation.

\subsection{Stale Placement Views}\label{stale-placement-views}

 Each Dispatch replica retains a recent snapshot of the Pools it serves. Requests handled between two refreshes share much of the same placement information.

Strict least-load selection becomes correlated under this access pattern. Let \(P\) be a Pool and let \(\ell_{\mathrm{snap}}(w)\) denote the cached load of Worker \(w\). A replica using Strict-Min selects

\begin{equation}\label{eq:6}
w^{*} = \arg\min_{w \in P}\ell_{\mathrm{snap}}(w).
\end{equation}

Suppose several Dispatch replicas observe the same \(w^{*}\) before their snapshots refresh. Assignments made after that snapshot are absent from
other replicas' views, so concurrent requests repeatedly target the same
cached minimum. The resulting concentration increases load skew and
failed admission attempts.

DHSched combines replica-local corrections with randomized selection among lightly loaded candidates (Figure 4). 

Capacity is then
checked against authoritative shared state before the assignment commits.

\begin{figure*}[tbp]
\centering
\includegraphics[width=\linewidth]{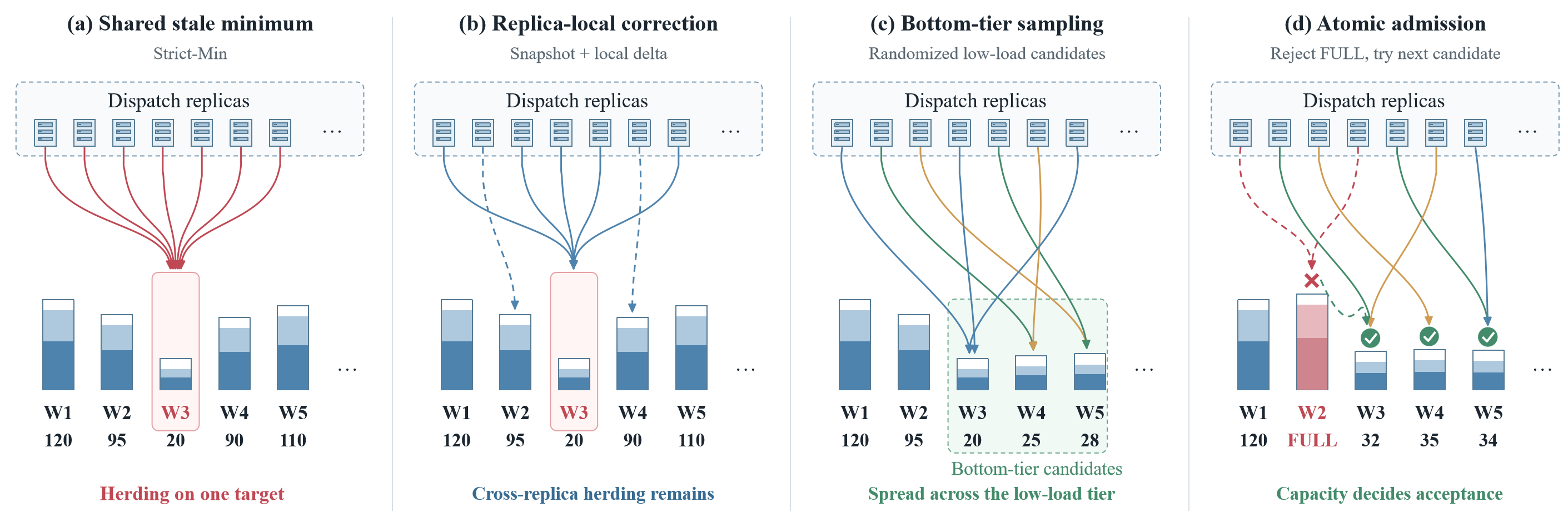}
\caption{Placement under stale views: shared-minimum herding, replica-local correction, bottom-tier sampling, and authoritative capacity admission.}\label{fig:6}
\end{figure*}

\subsection{Replica-Local Scheduling View}\label{replica-local-scheduling-view}

For replica \(r\), let \(\ell_{\mathrm{snap}}(w)\) denote the load of Worker \(w\) in the latest cached snapshot. Let \(\delta_{r}(w)\) denote assignments to \(w\) that replica \(r\) has committed since that snapshot.

DHSched ranks Workers using the effective load

\begin{equation}\label{eq:7}
{\widehat{\ell}}_{r}(w) = \ell_{\mathrm{snap}}(w) + \delta_{r}(w).
\end{equation}

After a successful assignment, the local delta increments immediately, so the replica\textquotesingle s next placement includes that assignment before the shared view refreshes.

When a newer snapshot incorporates these assignments, the corresponding local delta is rebased. The same accounting is aggregated at the Pool level. For Pool \(P\),

\begin{equation}\label{eq:8}
\Delta_{r}(P) = \sum_{w \in P}^{}\delta_{r}(w).
\end{equation}

The local delta covers assignments issued by replica \(r\). Assignments from other replicas remain absent until the shared view refreshes. DHSched combines this local correction with randomized Worker selection inside each Pool.

Local deltas necessitate no Dispatch-to-Dispatch communication or shared write on the ranking path. 

\subsection{Cross-Pool Selection}\label{cross-pool-selection}

A DigitalID can be served by more than one Pool. Eligible Pools differ in Worker count, total capacity, GPU class, region, and configured deployment priority. Raw live counts are not comparable across Pools with different capacities.

For Pool \(P\), define the cached aggregate load

\begin{equation}\label{eq:9}
L_{\mathrm{snap}}(P) = \sum_{w \in P}^{}\ell_{\mathrm{snap}}(w)
\end{equation}

and the total configured capacity

\begin{equation}\label{eq:10}
C(P) = \sum_{w \in P}^{}C_{w},
\end{equation}

where \(C_{w}\) is the capacity of Worker \(w\).

Replica \(r\) computes the Pool waterline as

\begin{equation}\label{eq:11}
\mathrm{WL}_{r}(P) = \frac{L_{\mathrm{snap}}(P) + \Delta_{r}(P)}{C(P)}.
\end{equation}

The waterline combines cached and replica-local load and normalizes by Pool capacity. 

Dispatch first filters Pools by DigitalID eligibility and orders them by configured deployment priority. Among eligible Pools with equal priority, it prefers the Pool with the smaller \(\mathrm{WL}_{r}(P)\).

\begin{figure}[t]
\centering
\includegraphics[width=\columnwidth]{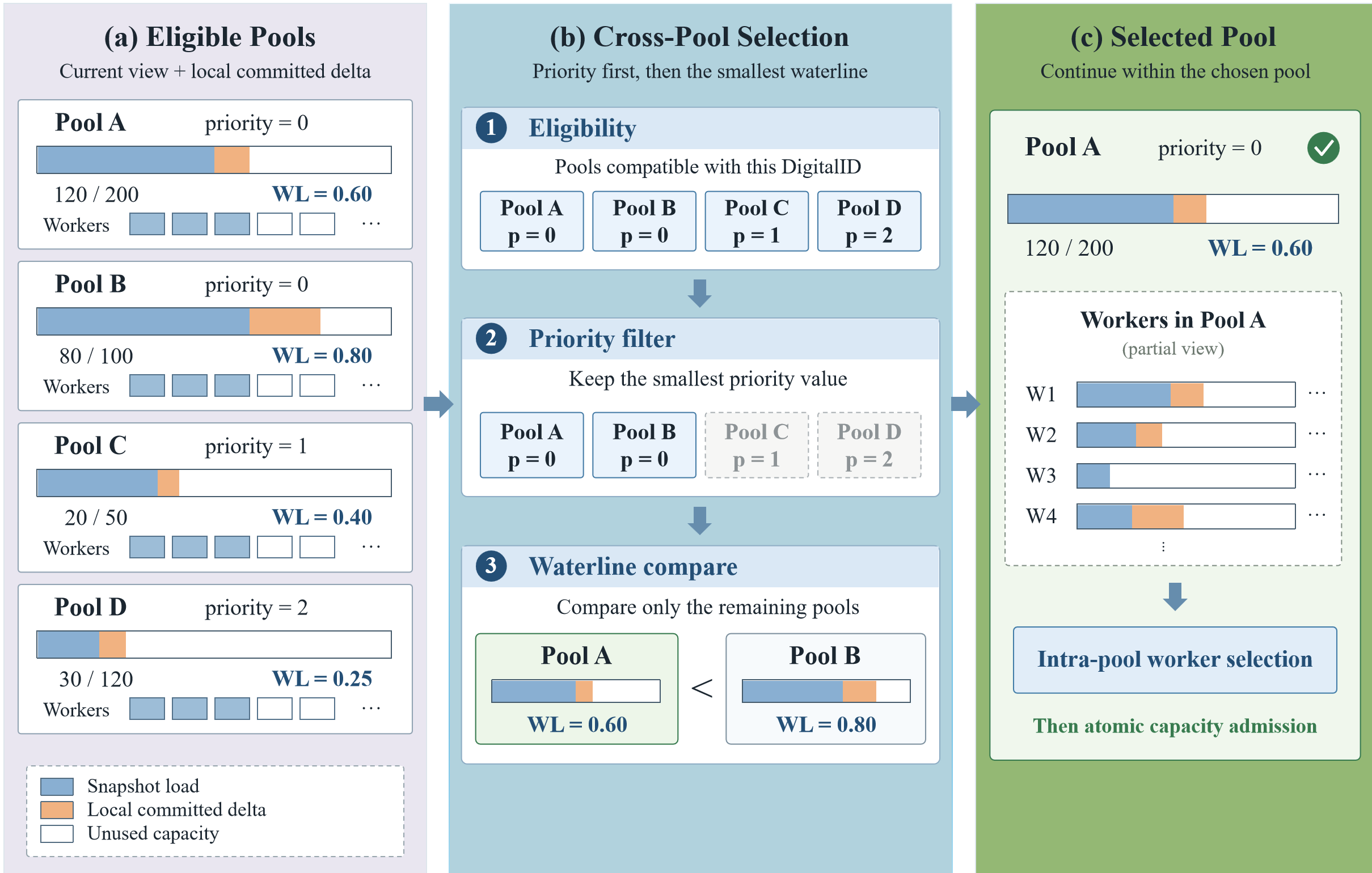}
\caption{Cross-Pool selection applies eligibility and deployment priority before comparing capacity-normalized waterlines.}
\label{fig:7}
\end{figure}

 Once a Pool has been chosen, Dispatch selects a Worker from that Pool using the policy described next. If the remaining candidates in one Pool fail admission, placement proceeds to the next eligible Pool.

\subsection{Bottom-Tier Worker Selection}\label{bottom-tier-worker-selection}

Within a selected Pool, Dispatch ranks Workers by their effective load \({\widehat{\ell}}_{r}(w)\). Picking the minimum directly recreates the correlated behavior from Section~\ref{stale-placement-views}. DHSched instead samples from a low-load tier.

Let

\begin{equation}\label{eq:12}
\ell_{\min} = \min_{w \in P}{\widehat{\ell}}_{r}(w)
\end{equation}

and let \(k(P)\) denote the configured tier size for Pool \(P\). The implementation keeps a minimum tier size and expands the tier with Pool size. It also limits the tier to Workers whose effective load remains within a configured score window \(\tau\) of the current minimum.

The candidate set is

\begin{equation}\label{eq:13}
\begin{aligned}B_r(P)=\bigl\{w\in P\bigm|{}&\operatorname{rank}_r(w)\leq k(P)\\&\land\;\widehat{\ell}_r(w)\leq\ell_{\min}+\tau\bigr\}.\end{aligned}
\end{equation}

Here, \(\operatorname{rank}_{r}(w)\) is the rank of \(w\) under the ordering induced by \({\widehat{\ell}}_{r}\).

Dispatch samples candidates uniformly without replacement from \(B_{r}(P)\). For a candidate \(w\),

\begin{equation}\label{eq:14}
w \sim \operatorname{Uniform}\left( B_{r}(P) \right).
\end{equation}

Each sampled Worker is passed to capacity admission. A FULL response advances the request to another sampled candidate until the attempt budget is exhausted.

The tier keeps selection near the low-load end of the Pool while removing the cached minimum as a common target. The local delta shifts Workers selected recently by the same replica upward in the ranking. Random sampling distributes choices that originate from different replicas and still share similar snapshots.

\subsection{Atomic Capacity Admission}\label{atomic-capacity-admission}

For candidate \(w\), Dispatch executes an atomic store-side reservation. Its logical form is given in Algorithm 3.

\begin{center}
\vspace{2pt}
\includegraphics[width=0.98\columnwidth]{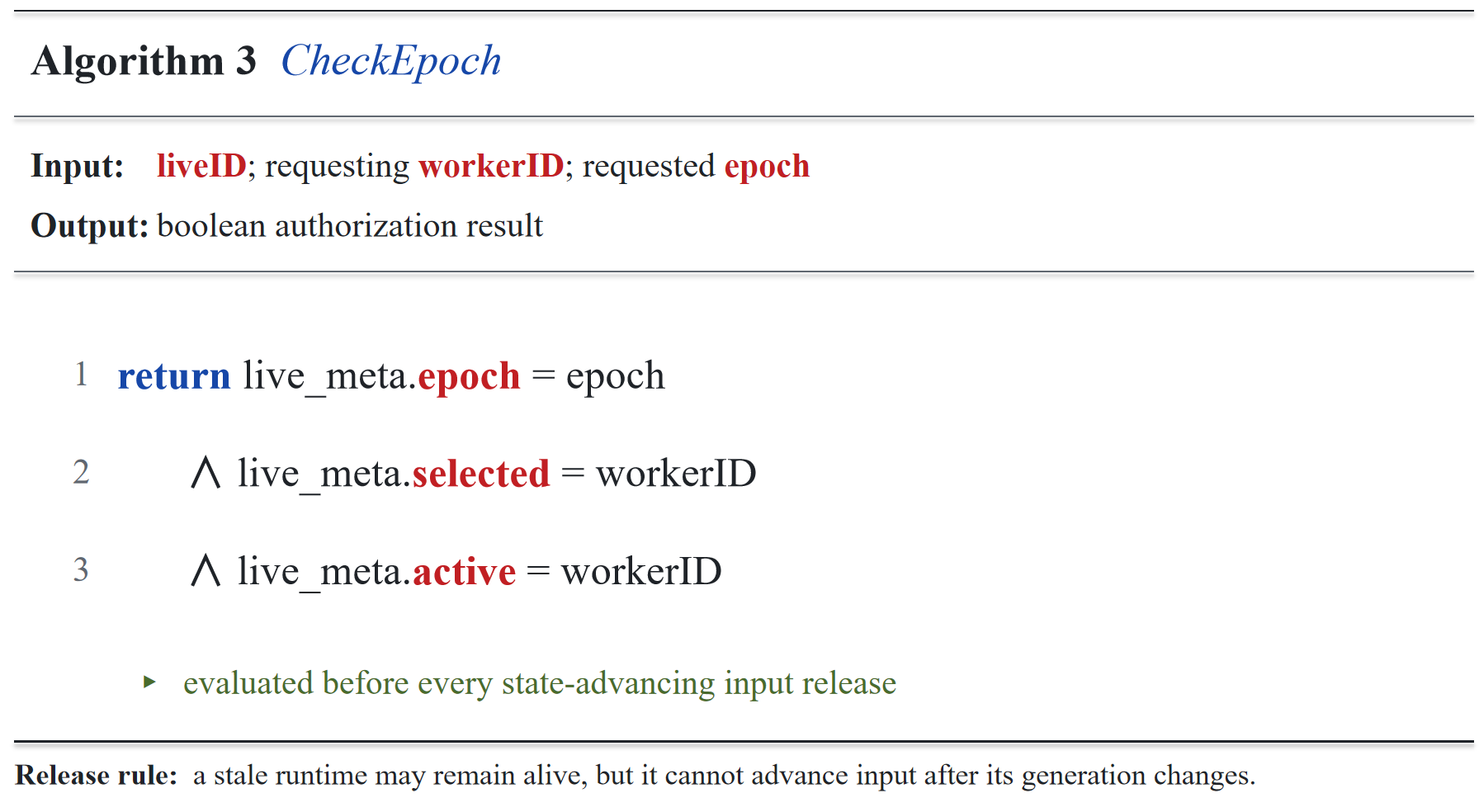}
\vspace{2pt}
\end{center}

For every Worker \(w\), the admission path maintains

\begin{equation}\label{eq:15}
\mathsf{committed}(w) + \mathsf{reserved}(w) \leq C_{w}.
\end{equation}

Concurrent reservations for the same Worker enter a single order at the store. Once the available slots have been consumed, later attempts observe the updated membership and return FULL.

A successful generation commit converts the corresponding reservation into committed membership. Cancellation and cleanup remove a reservation only when its reservation identifier and generation still match the state created by that operation.

A stale snapshot sometimes sends Dispatch to a Worker that has already filled. The admission step rejects that candidate and placement continues elsewhere. The cached scheduling view determines candidate order. Authoritative Worker membership determines whether the assignment is accepted.

Dispatch replicas may rank from different snapshots while the atomic reservation enforces the same capacity bound.

%% file: sections/06-source-independent-ownership-transfer.tex
\section{Source-Independent Ownership Transfer}\label{source-independent-ownership-transfer}

Recovery uses the same generation transition as normal assignment. The difference is that the current Worker may already be unavailable when the transition begins. DHSched records the generation that triggered recovery and accepts recovery work only while that generation remains current. After a replacement generation commits, the target reconstructs the session from shared state and takes over the persistent input path.

\subsection{Epoch-Guarded Recovery}\label{epoch-guarded-recovery}

Worker failure and session stalls generate recovery work asynchronously. For each affected live, DHSched records the liveID together with the epoch observed at detection time.

Let a recovery item for live \(s\) carry source epoch \(e_{s}\). Before acting on the item, a recovery consumer reads the current session generation. Recovery proceeds only when

\begin{equation}\label{eq:17}
\mathsf{epoch}(s) = e_{s}.
\end{equation}

If the current epoch is higher, another create, reassignment, or recovery has already advanced the live. The item is then discarded without modifying ownership.

For a current item, Dispatch selects an eligible Pool, admits a replacement Worker, and commits a new generation against \(e_{s}\). A successful commit advances the epoch and clears the previous data-plane claim. The replacement Worker then creates its runtime and claims the new generation through Infer-Controller.

The recovery queue provides at-least-once processing, so the same item may be delivered more than once. The source-epoch check makes these retries harmless. Once any operation advances the live beyond \(e_{s}\), every delayed recovery item carrying \(e_{s}\) becomes stale.

A Worker that resumes after recovery is handled in the same way. Its requests still carry the earlier generation and fail the generation check before new input is released.

\subsection{Source-Independent Reconstruction}\label{source-independent-reconstruction}

After the replacement generation commits, the target reads the session metadata and pending input from shared state. It creates the local avatar runtime, establishes the RTC connection, and issues its first PullText for the new generation. Infer-Controller completes the claim before releasing pending input.

The source Worker sends no GPU state, renderer state, or RTC runtime to the target. Recovery continues when the source has crashed or remains partitioned from the rest of the system.

Target-side runtime creation and RTC establishment still contribute to recovery time, but neither step requires the previous runtime to respond.

\subsection{Stream-Aware Planned Drain}\label{stream-aware-planned-drain}

\begin{figure*}[tbp]
\centering
\includegraphics[width=\linewidth]{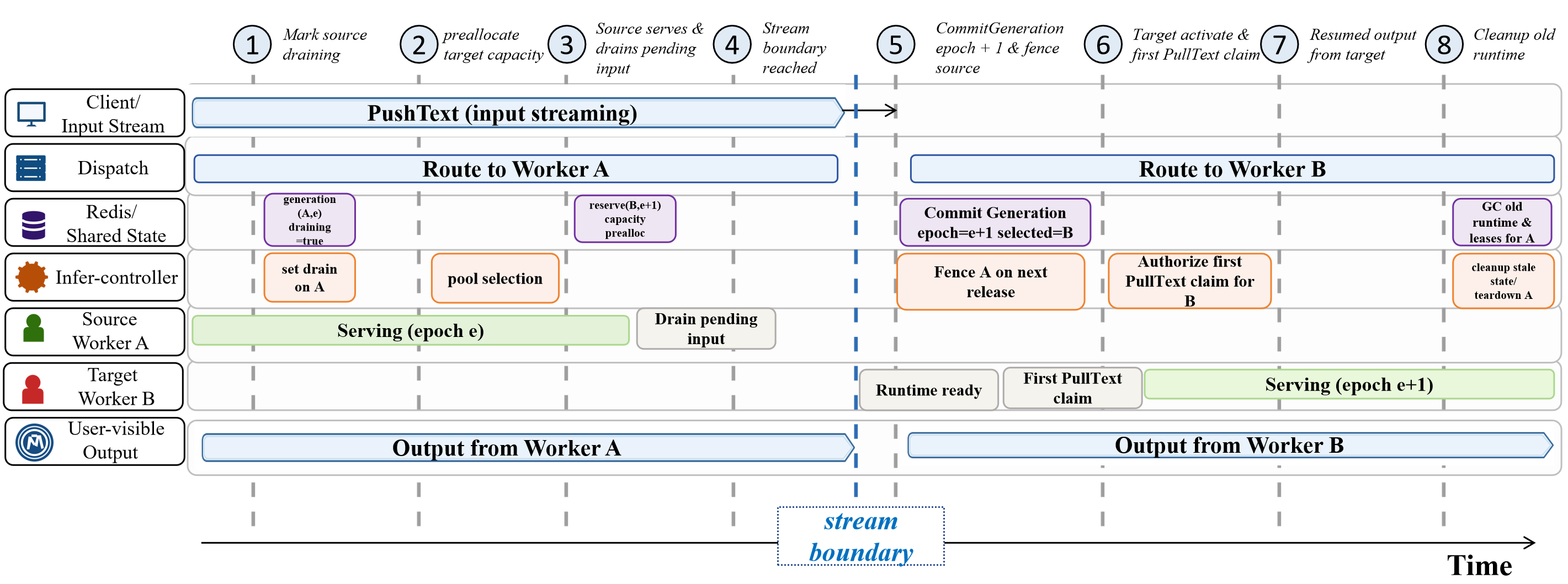}
\caption{Planned drain prepares replacement capacity, commits at a stream boundary, fences the source, and activates the target before cleanup.}\label{fig:8}
\end{figure*}

Planned drain starts while the source Worker remains healthy. DHSched keeps the current generation serving while it reserves replacement capacity and waits for the configured stream boundary (Figure~\ref{fig:8}).

At the boundary, Dispatch commits the replacement generation. Subsequent generation checks fence the source, and the target claims the new generation through its first PullText. DHSched limits concurrent cutovers from one Worker so that runtime creation and RTC joins do not arrive as one burst. A bounded wait prevents continuously active streams from delaying drain indefinitely.

%% file: sections/07-implementation.tex
\section{Implementation}\label{implementation}

DHSched is implemented in our production avatar-serving stack using Redis-backed authoritative metadata, per-replica placement caches, and Worker-side reconciliation. Control-path updates commit shared metadata, while GPU and RTC runtime creation, teardown, and cleanup proceed asynchronously.

\subsection{Shared State and Atomic Updates}\label{shared-state-and-atomic-updates}

Table~\ref{tab:state} summarizes the implementation state. Ownership and capacity records are authoritative, while Pool views and Worker-local state support placement, activation, and cleanup. Operations that span several records for the same live or Worker use short-lived entity locks around their multi-step work. Capacity reservations and generation transitions use atomic Lua or compare-and-set operations. The lock limits overlap among surrounding steps, while the conditional write determines whether the state transition commits. Dispatch keeps no durable ownership state locally. If a replica fails after a successful write, a retry reconstructs the committed result from Redis.

\begin{table*}[tbp]
\caption{Implementation state and its role.}\label{tab:state}
\centering\small
\renewcommand{\arraystretch}{1.12}
\begin{tabular}{@{}>{\raggedright\arraybackslash}p{\dimexpr 0.22\linewidth-1.3333333333333333\tabcolsep\relax} >{\raggedright\arraybackslash}p{\dimexpr 0.24\linewidth-1.3333333333333333\tabcolsep\relax} >{\raggedright\arraybackslash}p{\dimexpr 0.54\linewidth-1.3333333333333333\tabcolsep\relax}@{}}
\toprule
\textbf{State} & \textbf{Location} & \textbf{Role} \\
\midrule
live\_meta & Authoritative shared store & Current epoch, selected Worker, active claim, room metadata, lifecycle state \\
worker\_weight & Authoritative shared store & Per-Worker committed and reserved live membership for admission \\
pool\_* views & Shared store and Dispatch cache & Pool membership, load, capacity, priority, and waterline inputs \\
PushText stream & Infer-Controller / shared store & Pending input, request identifiers, acknowledgements, and replay position \\
migration records & Shared store & Epoch-tagged recovery work and retry lease \\
avatar runtime and RTC connection & Worker & GPU execution, rendering context, and media connection \\
recent request IDs & Worker & Bounded duplicate suppression across delivery retries \\
\bottomrule
\end{tabular}
\end{table*}

\subsection{Create and Activation Path}\label{create-and-activation-path}

CreateLive executes the placement and admission path from Section~\ref{capacity-safe-placement-of-new-generations}, commits the resulting generation, converts its reservation into committed membership, and publishes the live in the target Worker\textquotesingle s desired set. The control request returns after the shared-state update. Runtime activation proceeds asynchronously. The Worker reconciles its desired-live set against the runtimes present on the machine, creating the avatar runtime and establishing the RTC connection for newly assigned lives. The Worker then follows the claim and per-input validation path from Section~\ref{cross-plane-ownership-architecture}.

\subsection{Replica-Local Placement State}\label{replica-local-placement-state}

Each Dispatch replica keeps an in-memory snapshot of Pool membership, Worker load, and capacity. Each snapshot uses a refresh TTL of roughly 200 ms to 1 s. Pool waterlines, effective Worker scores, and bottom-tier candidates are computed from this cached view.

The replica also maintains the local deltas defined in Section~\ref{capacity-safe-placement-of-new-generations}. A delta is incremented only after the corresponding assignment commits. When a newer shared snapshot incorporates those assignments, the replica rebases the local value against the refreshed load.

Candidate ranking normally runs from the in-memory view. Redis is accessed again for authoritative capacity admission and generation commit. With \(N\) candidate attempts, the current CreateLive path issues approximately \(8 + 6N\) Redis operations. The common case of two to three attempts uses about twenty operations.

\subsection{Worker Reconciliation and Input Delivery}\label{worker-reconciliation-and-input-delivery}

Workers periodically report resource usage, live membership, and runtime state and reconcile their desired-live sets against the runtimes present locally. Missing desired lives are created, while runtimes absent from the desired set are removed.

Dispatch completes the control update without waiting for GPU runtime construction or RTC establishment. Infer-Controller applies the claim and generation checks from Section~\ref{cross-plane-ownership-architecture} on the input path. A failed validation stops further consumption and triggers teardown of the obsolete runtime.

Input delivery retains at-least-once semantics. Workers keep a bounded history of recent request identifiers so that retries after lost acknowledgements do not repeatedly advance the local runtime.

\subsection{Failure Recovery}\label{failure-recovery}

Worker reports feed a liveness monitor that scans once per second. When a Worker expires, Dispatch removes it from new placement and enqueues its affected sessions for recovery. Each migration record carries the liveID and source epoch observed when the work was created.

Recovery consumers process these records asynchronously under a retry lease. Before entering the recovery path in Section~\ref{source-independent-ownership-transfer}, a consumer compares the recorded source epoch with the current generation and discards superseded records. Current work follows normal placement and reconstructs the replacement from shared metadata and pending input after the new generation commits.

\subsection{Planned Drain and Background Cleanup}\label{planned-drain-and-background-cleanup}

The implementation limits concurrent cutovers from one Worker. A bounded wait, currently 60 s, prevents continuously active streams from delaying drain indefinitely.

Session lifecycle updates span authoritative records, derived membership and reservation state, and Worker-local runtimes. A crash or lost response can leave stale reservations, migration records, derived membership, or local runtimes after ownership has advanced.

Background reconciliation repairs this state. Worker reports refresh live membership and resource state, while Workers reconcile local runtimes against their desired-live sets. Cleanup removes reservations and migration records only when the stored epoch or reservation identifier still matches the operation that created them. Placement traffic also corrects stale load metadata when authoritative Worker membership disagrees with the cached derived value.

%% file: sections/08-evaluation.tex
\section{Evaluation}\label{evaluation}

We first test execution authority under concurrent ownership changes,
followed by placement, capacity admission, and recovery. Production
observations provide the operating context for these controlled experiments.

\subsection{Experimental Methodology}\label{experimental-methodology}

The placement experiments use 128 Workers, a 16,000-session burst,
16 Dispatch replicas by default, a three-candidate attempt budget,
and 30 common seeds. Every policy receives the same requests, initial
loads, Worker capacities, eligibility constraints, snapshot-refresh
schedule, retry budget, and authoritative atomic admission. The snapshot
sweep ranges from 1 to 4,000 placements per refresh; the replica sweep
ranges from 1 to 32. The hierarchy experiment uses eight heterogeneous
Pools and twelve DigitalID classes eligible for two to eight Pools.
Eighty percent of demand follows the free-capacity share of each class's
preferred Pool tier; the remaining 20\% is flexible traffic.

For homogeneous placement, load spread is the difference between the
maximum and minimum assigned-session counts across Workers. Under
heterogeneous capacity, normalized-utilization spread is the corresponding
difference after dividing each Worker's load by its capacity. A rejected
candidate is an admission attempt that returns FULL. A failed Create
exhausts its candidate budget without committing an assignment. Off-tier
placement is the fraction of successful sessions assigned outside the
DigitalID's preferred Pool tier.

The authority experiment executes 100,000 randomized runs per protocol.
Each run contains a reassignment, delayed pulls from the previous Worker,
pulls from the target, duplicate control work, a delayed recovery item,
and---with probability 0.55---a return to the same Worker at a higher
epoch. A run violates single-owner execution if state-advancing input is
released to more than one ownership generation. A resurrection occurs
when delayed recovery work reinstalls an ownership generation after a
newer generation has committed. Operation types and delays are sampled
independently from the workload distributions used throughout the
authority experiments. Capacity admission uses 20,000 trials at each of
2--64 concurrent contenders. Recovery uses 20,000 trials at each
delayed-item probability from 0.01 to 0.50.

We model three mechanisms described in related systems under the common
DHSched workload. The Orleans-AC model follows ActivationCount placement.
Each Dispatch replica ranks a random pair using the latest cached activation
count plus its own successful placements since that snapshot
\cite{orleansActivationCount}. Orleans-AC+Priority applies the configured
Pool priority before this choice and retains an eligible fallback. The
guard-lease model waits for lease recall or expiration before a changed
live resumes. The source-assisted model requires the current source to
remain reachable while runtime state is transferred to the target.

We also exercise the released Orleans 10.3.0 runtime. Placement uses its
unmodified ActivationCountBasedPlacement policy across 2--16 local silos,
twenty fresh-process runs per configuration, 16,000 activations per run,
and 16 concurrent issuers. The failure experiment starts 4--16 silos,
activates 16,000 grains, and compares graceful leave with abrupt termination
under the default and preview distributed directories. After a non-primary
silo leaves, the client continuously invokes up to 300 grains that had
been hosted there. Each failure configuration uses five fresh-process
runs and a 90 s recovery window.

\subsection{Production Context}\label{production-context}

\afterpage{%
\begin{table*}[!tbp]
\caption{Production operating context on 10 June 2026.}
\label{tab:production}
\centering
\small
\renewcommand{\arraystretch}{1.12}
\begin{tabular}{
@{}
>{\raggedright\arraybackslash}p{\dimexpr 0.24\linewidth-1.0\tabcolsep\relax}
>{\raggedright\arraybackslash}p{\dimexpr 0.76\linewidth-1.0\tabcolsep\relax}
@{}
}
\toprule
\textbf{Production metric} &
\textbf{Observed value (10 June 2026)} \\
\midrule
Peak concurrent sessions &
49,987 \\

Daily control/data operations &
125,728 CreateLive; 18.47M PushText \\

Tail latency &
98.3 ms CreateLive P99; 9.9 ms WorkerSelect P99;
16.5 ms PushText P99 \\

Ownership and recovery &
9,927 completed session migrations; 0 observed dual-owner;
0 observed overshoot; 51 Worker-offline detections;
2,040 Worker-level migrations \\
\bottomrule
\end{tabular}
\end{table*}
}

Table~\ref{tab:production} summarizes DHSched's production operating
context. Table~\ref{tab:published} places these observations alongside
representative systems using each system's native scheduling unit and
published operating evidence.

\afterpage{%
\begin{table*}[!tbp]
\caption{Published-system context in each system's native scheduling unit.}
\label{tab:published}
\centering
\small
\renewcommand{\arraystretch}{1.12}
\begin{tabular}{
@{}
>{\raggedright\arraybackslash}p{\dimexpr 0.12\linewidth-1.5\tabcolsep\relax}
>{\raggedright\arraybackslash}p{\dimexpr 0.18\linewidth-1.5\tabcolsep\relax}
>{\raggedright\arraybackslash}p{\dimexpr 0.40\linewidth-1.5\tabcolsep\relax}
>{\raggedright\arraybackslash}p{\dimexpr 0.30\linewidth-1.5\tabcolsep\relax}
@{}
}
\toprule
\textbf{System} &
\textbf{Scheduling unit} &
\textbf{Published operating context} &
\textbf{Ownership / transfer path} \\
\midrule

DHSched &
Long-lived avatar live &
49,987 peak concurrent production lives; 9,927 session migrations &
Generation fencing; source-independent reconstruction \\

TurboServe &
Long-lived streaming-video session &
Production traces; up to 616.8 average active sessions in a reported
interval; clusters up to 64 B300 GPUs &
GPU-CPU offload; NCCL GPU-GPU migration \\

Slicer &
Shard / key &
2--6M req/s production traffic; optional strong-consistency feature
implemented but not yet production-deployed at publication &
Authoritative assignment; lease-coordinated strong consistency \\

Llumnix &
In-flight LLM request &
OSDI'24 multi-instance serving evaluation with runtime rescheduling &
Source-assisted KV-cache live migration \\
\bottomrule
\end{tabular}
\end{table*}
}

\subsection{Per-Input Execution Authority}
\label{per-input-execution-authority}

Assignment-only and Epoch-at-commit violate single-owner execution in
99.9\% of randomized runs; Claim-only still violates it in 99.4\%.
Per-release WorkerID-only fencing reduces the rate to 14.0\%, but cannot
distinguish an old A runtime from a new A generation after A-to-B-to-A
reassignment. In that subset, 25.5\% of runs violate single-owner.
DHSched records no dual-owner or stale-generation release. The A-to-B-to-A
cases isolate the role of the epoch, since successive generations can
select the same Worker.

\begin{figure}[!htbp]
\centering
\includegraphics[width=\columnwidth]{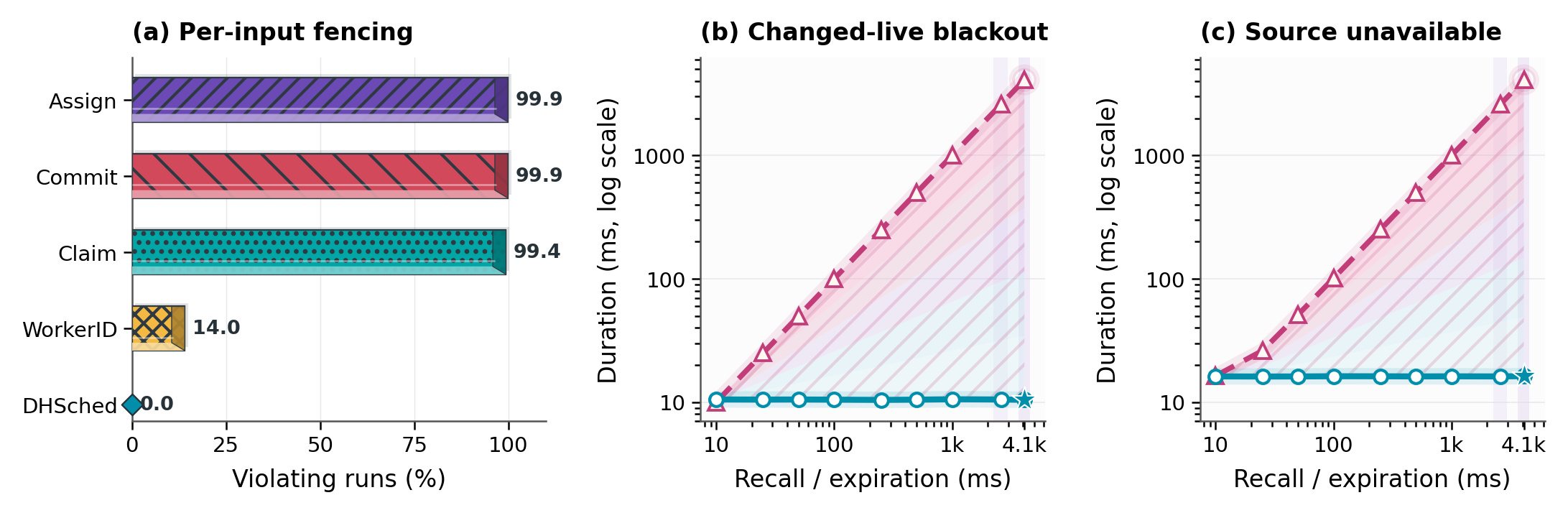}
\caption{Execution authority. (a) Only the full per-input generation guard
prevents all tested violations. (b--c) Guard-lease blackout and takeover
follow recall or expiration; both protocols prevent dual ownership.
Lower is better.}
\label{fig:authority}
\end{figure}

The guard-lease model and DHSched both record zero dual-owner violations.
Slicer reports its optional strong-consistency assignment feature as
implemented but not yet deployed by production customers at publication
\cite{adya2016slicer}. In DHSched's production deployment, the single-owner
path covered 9,927 session migrations on a day that reached 49,987
concurrent lives, with no observed dual-owner conflict
(Table~\ref{tab:production}). Recall or expiration makes a changed live
unavailable for a duration determined by the lease path. A DHSched
generation commit revokes the previous generation's input authority,
while target activation proceeds independently of source liveness and
obtains authority only after Claim. Figure~\ref{fig:authority} marks
Slicer's published 2.6 s median and 4.1 s P99 recall periods as reference
points.

\textbf{Executable Redis-backed validation.} The results above exercise a model
of the protocol. We additionally implemented Commit, Claim, per-input
validation, idempotent retry, and Reserve as Redis Lua operations driven by
independent TCP clients, which is a reference implementation of the published
protocol, not the production code path. Across 10{,}000
A$\rightarrow$B$\rightarrow$A schedules, the generation guard releases stale
input in none of them, and WorkerID-only validation does so in 20.2\%. A real
store reproduces the ordering the evaluator reports at 25.5\%. Across
10{,}000 post-commit socket failures, a retry returns the committed generation
in every run, with no second successor and no duplicate charge
(Table~\ref{tab:redis}).

\begin{table}[!t]
\caption{Executable Redis-backed validation against a real Redis store.
Lower is better in the first two rows, higher in the last three.}
\label{tab:redis}
\centering
\small
\renewcommand{\arraystretch}{1.12}
\begin{tabular}{@{}llr@{}}
\toprule
\textbf{Check} & \textbf{Guard} & \textbf{Result} \\
\midrule
Stale release, A$\rightarrow$B$\rightarrow$A & WorkerID-only & 20.2\% \\
 & Generation & 0.0\% \\
\addlinespace
Retry after lost response & Same generation & 10{,}000/10{,}000 \\
 & No second successor & 10{,}000/10{,}000 \\
 & No duplicate charge & 10{,}000/10{,}000 \\
\bottomrule
\end{tabular}
\end{table}

\subsection{Placement under Stale and Hierarchical Views}
\label{placement-under-stale-and-hierarchical-views}

At 1,000 placements per snapshot, Strict-Min reaches 205.7 max-min spread
and 27,549 rejected candidates. The Orleans-AC model reaches 66.3 spread
and DHSched 63.6.

The hierarchy workload requires placement to respect Pool preference while
completing the burst under the same candidate budget. The Orleans-AC model
places 45.4\% of successful sessions outside the preferred Pool tier and
averages 609.9 failed Creates. Adding Pool priority reduces its off-tier
rate to 14.1\%, but 365.3 Creates still fail on average. Cached Least-Load
comes close to finishing the burst and still sends 84.0\% of sessions
outside the preferred tier. DHSched reaches
12.3\% off-tier placement and completes all 16,000 Creates.

\begin{figure}[!htbp]
\centering
\includegraphics[width=\columnwidth]{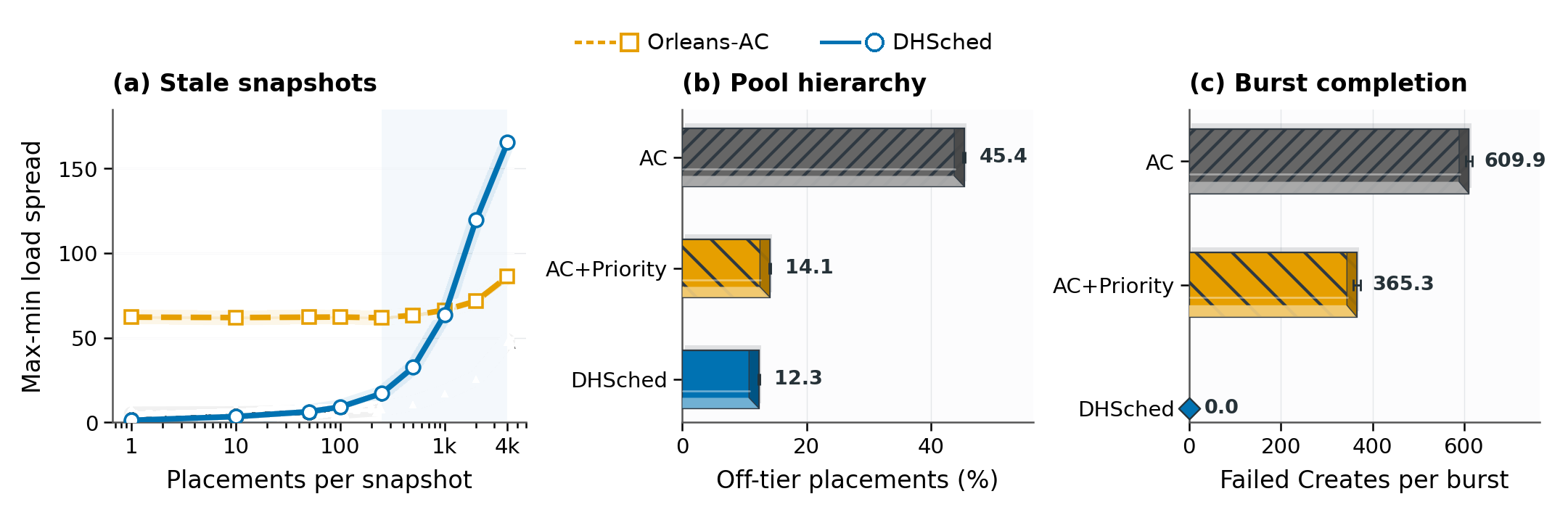}
\caption{Stale and hierarchical placement. (a) Orleans-AC and DHSched are
comparable under stale counts. (b--c) Priority
adaptation reduces off-tier placement; DHSched completes all Creates.
Means and 95\% CIs over 30 common seeds; lower is better.}
\label{fig:placement}
\end{figure}

\begin{figure}[!htbp]
\centering
\includegraphics[width=\columnwidth]{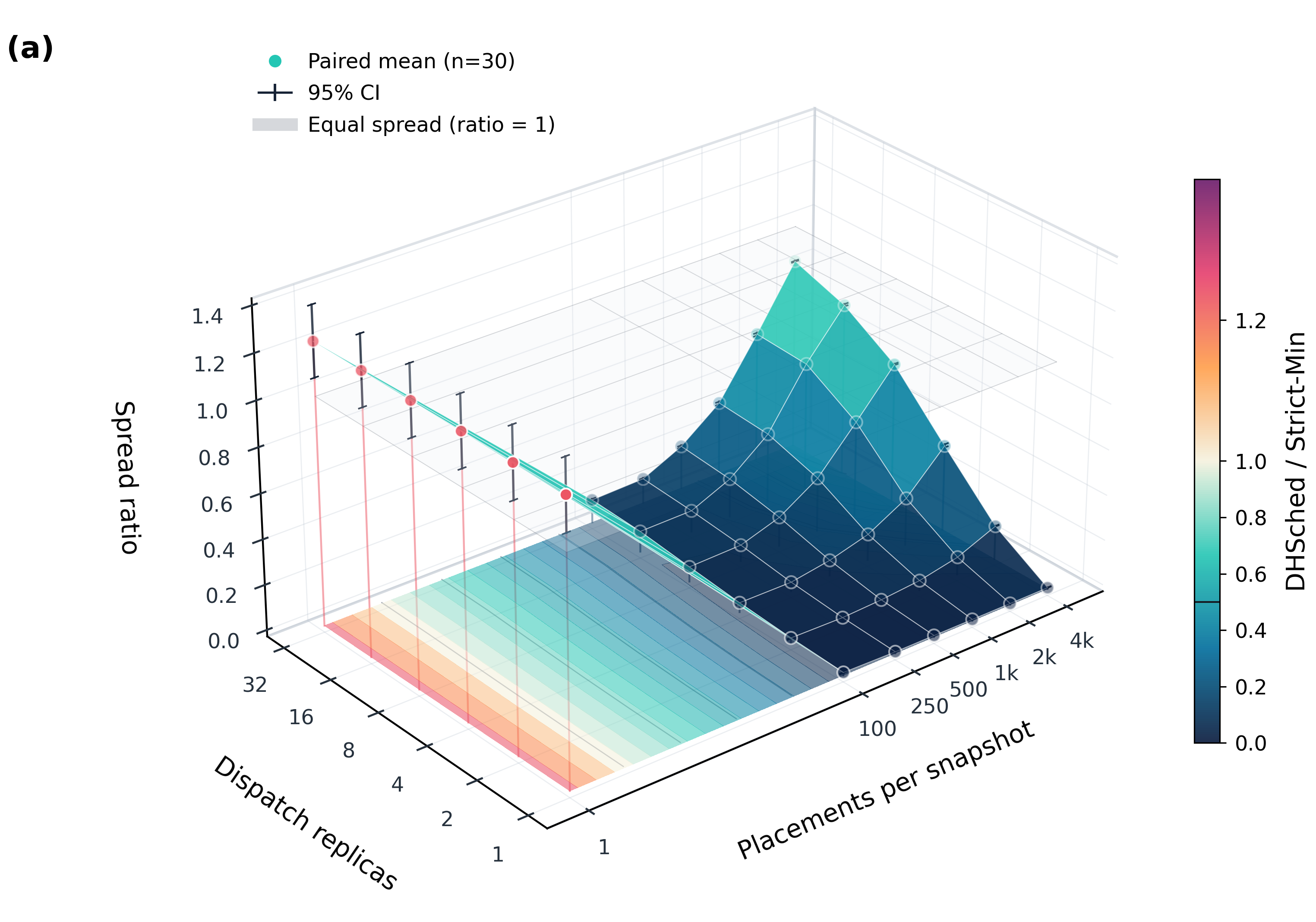}
\caption{Joint sensitivity to snapshot staleness and Dispatch parallelism.
Paired DHSched/Strict-Min spread ratios over 30 common seeds, with 95\%
CIs; the plane marks equal spread.}
\label{fig:sensitivity}
\end{figure}

\textbf{Joint sensitivity.}
We next vary snapshot staleness and Dispatch parallelism together.
Figure~\ref{fig:sensitivity} reports the paired ratio of DHSched to
Strict-Min max-min spread across 30 common seeds. With fresh placement
views, strict least-load remains competitive. As more placements share
the same cached snapshot, correlated choices increase its spread and
DHSched moves below the equal-spread boundary across the tested replica
counts. Increasing the number of Dispatch replicas weakens the reach of
replica-local correction because each replica observes only its own recent
assignments.

\textbf{Released Orleans runtime.}
We additionally run Orleans 10.3.0 with its unmodified
ActivationCountBasedPlacement policy. Across 2--16 local silos and twenty
independent 16,000-activation bursts per configuration, all 1.28 million
requested activations complete. Mean max-min spread ranges from 3.3 to
11.0 activations, or 0.041\%--1.10\% of mean silo load. Comparing the
runtime means with the model sweep places the local runtime near the
model's tens-to-roughly-one-hundred placements-per-snapshot region.
The 1,000-placement point evaluates a substantially staler view.

\begin{figure}[!htbp]
\centering
\includegraphics[width=\columnwidth]{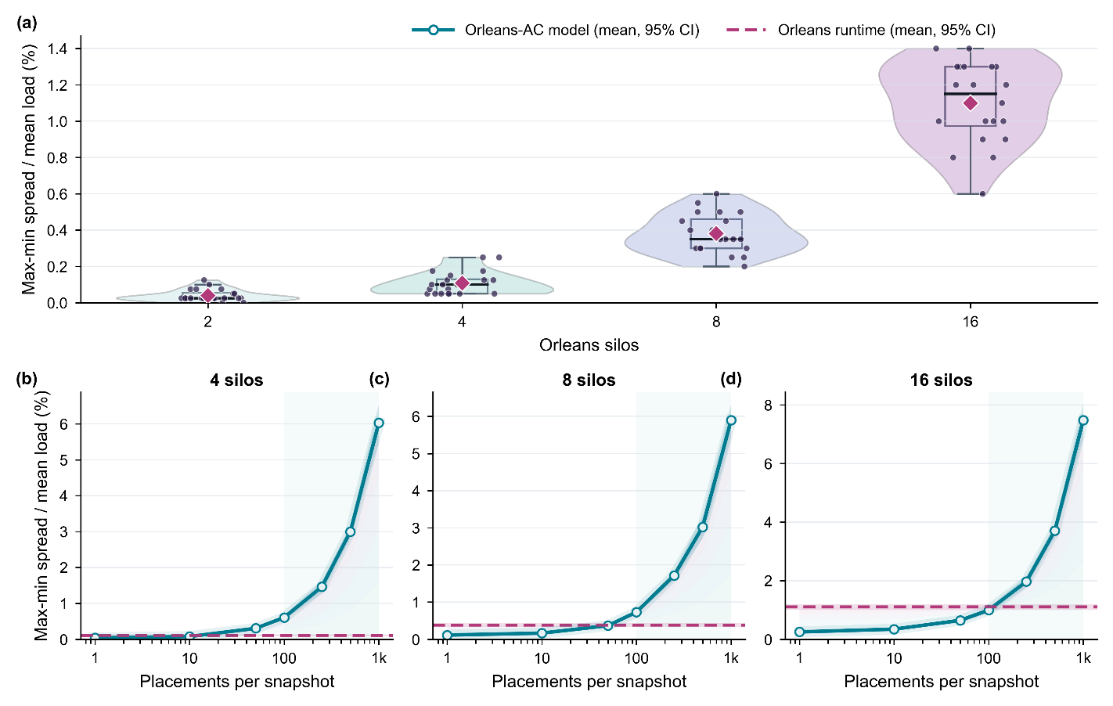}
\caption{Released Orleans placement. (a) Unmodified Orleans 10.3.0 across
2--16 local silos; boxplots summarize twenty independent 16,000-activation
bursts per configuration. (b--d) Runtime means align with model views aged
tens to roughly one hundred placements per snapshot.}
\label{fig:orleans-placement}
\end{figure}

\subsection{Atomic Capacity Admission}
\label{atomic-capacity-admission-1}

A non-atomic read-check-write overshoots capacity in 35.3\% of trials with
two contenders, 93.7\% with 16, and every trial with 64. Its maximum
overshoot grows from +1 to +62. Atomic Reserve records zero overshoot at
every contention level. Once the available slots are consumed, later
contenders receive FULL. The Redis reference implementation reproduces this
ordering against a real store: its weakened read-check-write path overshoots
in 54\% of two-client trials and in every trial from 16 clients onward, by as
many as 63 reservations at 64 clients, while Atomic Reserve stays at zero.

\begin{figure}[!htbp]
\centering
\includegraphics[width=\columnwidth]{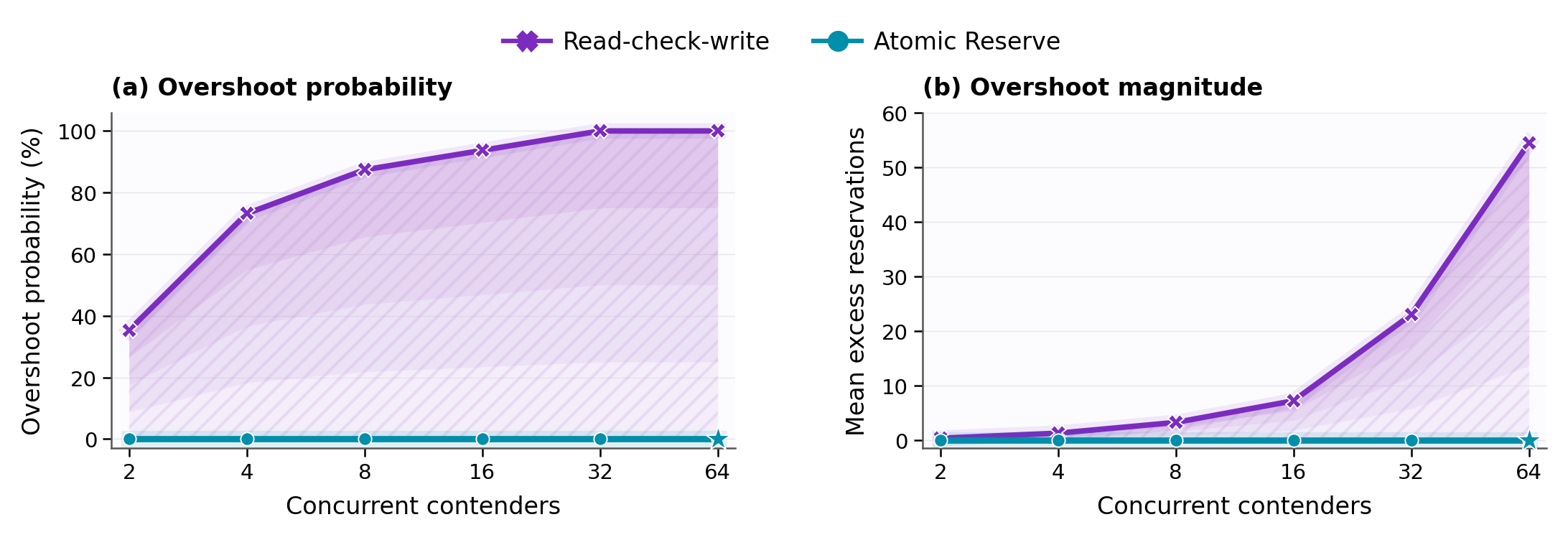}
\caption{Capacity admission under contention. Atomic Reserve preserves
the capacity bound.}
\label{fig:capacity}
\end{figure}

\subsection{Source-Independent Recovery}
\label{source-independent-recovery}

Without sourceEpoch validation, delayed recovery work resurrects stale
ownership in 6.8\% of runs at delayed-item probability 0.10 and 34.9\%
at probability 0.50. The guarded protocol records no stale write or
resurrection at every tested probability, discarding work after a newer
generation has committed.

Llumnix and TurboServe keep the source on the migration path by transferring
runtime state. DHSched reconstructs the target from shared metadata and
pending input.

Source-assisted completion falls with source availability, while DHSched
recovery remains independent of source participation.

\begin{figure}[!htbp]
\centering
\includegraphics[width=\columnwidth]{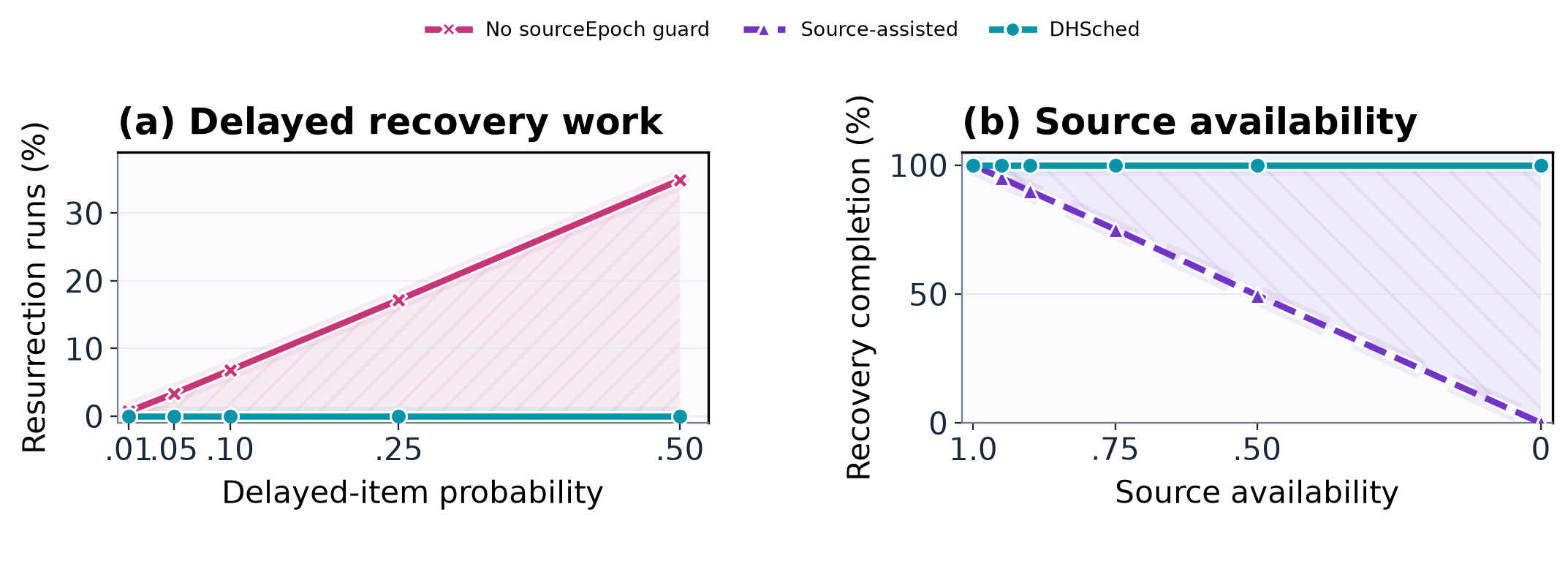}
\caption{Source-independent recovery. (a) Source-epoch guards reject stale
recovery work. (b) Source-assisted completion depends on source
availability, while DHSched recovery is independent of the source.
Lower is better in (a); higher in (b).}
\label{fig:source-recovery}
\end{figure}

\begin{figure}[!htbp]
\centering
\includegraphics[width=\columnwidth]{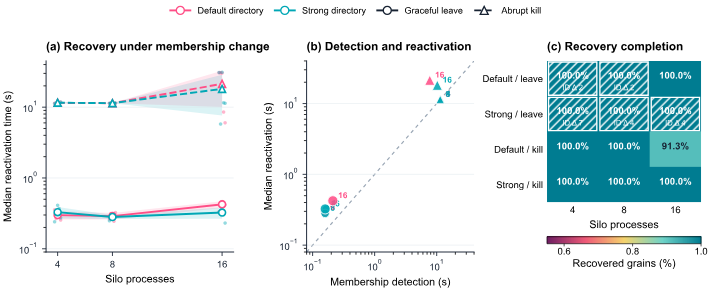}
\caption{Released Orleans recovery. (a--b) Graceful leave recovers in
hundreds of milliseconds; abrupt loss is detection-dominated at 4 and
8 silos. (c) At 16 silos, the default directory recovers 91.3\% within
90 s; the preview distributed directory recovers all targets.
Five fresh-process runs per configuration.}
\label{fig:orleans-recovery}
\end{figure}

\textbf{Released Orleans failure behavior.}
Across 4--16 localhost silos, graceful leave detects the membership change
in 156--212 ms on average and reaches median grain reactivation in
281--424 ms. Abrupt termination is detection-dominated at 4 and 8 silos,
with membership detection and median reactivation both near 11.5 s.
At 16 silos, the default directory recovers 91.3\% of targeted grains
within 90 s, while the preview distributed directory recovers all targeted
grains in these runs (Figure~\ref{fig:orleans-recovery}).

%% file: sections/10-discussion.tex
\section{Discussion}\label{discussion}

\textbf{Applicability.} DHSched applies to long-lived sessions whose recoverable state is externalized and whose state-advancing input crosses a boundary that can validate the current ownership generation. In our avatar stack, session metadata and pending input remain available while GPU and RTC runtimes are reconstructed after reassignment or failure. When recovery depends on execution state held only by the source runtime, handoff requires that state to be transferred.

\textbf{Shared-state boundary.}
The authoritative store orders ownership transitions and capacity admission, while placement snapshots, waterlines, and derived load state are refreshed asynchronously and affect only candidate ordering. Dispatch replicas retain no durable ownership state and can scale independently of the long-lived sessions they control. Scaling the authoritative store is a separate concern from scaling Dispatch.

\textbf{Recovery latency.} Source-independent recovery rebuilds the target from shared session metadata and pending input. It consists of target selection and admission, runtime creation, RTC establishment, and the first successful input claim. DHSched\textquotesingle s reconstruction path keeps ownership transfer available when source participation cannot be assumed.

%% file: sections/11-conclusion.tex
\section{Conclusion}\label{conclusion}

DHSched demonstrates that stateless peer controllers can manage long-lived stateful sessions by externalizing ownership and enforcing each generation on the persistent input path. Approximate load views guide placement while authoritative admission protects Worker capacity, and the same generation transition supports reassignment and recovery after source failure. In production, DHSched supported 49,987 concurrent sessions and 9,927 ownership transfers with no observed dual-owner conflicts or capacity overshoots. Stateful execution does not require a stateful scheduler when recoverable state is externalized and execution authority can be fenced independently.

%% file: refs.bib
@inproceedings{kwon2023vllm,
  author    = {Woosuk Kwon and Zhuohan Li and Siyuan Zhuang and Ying Sheng and
               Lianmin Zheng and Cody Hao Yu and Joseph E. Gonzalez and
               Hao Zhang and Ion Stoica},
  title     = {Efficient Memory Management for Large Language Model Serving
               with {PagedAttention}},
  booktitle = {Proceedings of the 29th Symposium on Operating Systems Principles
               (SOSP '23)},
  year      = {2023},
  publisher = {ACM}
}

@inproceedings{yu2022orca,
  author    = {Gyeong-In Yu and Joo Seong Jeong and Geon-Woo Kim and
               Soojeong Kim and Byung-Gon Chun},
  title     = {Orca: A Distributed Serving System for {Transformer-Based}
               Generative Models},
  booktitle = {16th USENIX Symposium on Operating Systems Design and
               Implementation (OSDI 22)},
  pages     = {521--538},
  year      = {2022},
  publisher = {USENIX Association}
}

@inproceedings{li2023alpaserve,
  author    = {Zhuohan Li and Lianmin Zheng and Yinmin Zhong and Vincent Liu and
               Ying Sheng and Xin Jin and Yanping Huang and Zhifeng Chen and
               Hao Zhang and Joseph E. Gonzalez and Ion Stoica},
  title     = {{AlpaServe}: Statistical Multiplexing with Model Parallelism
               for Deep Learning Serving},
  booktitle = {17th USENIX Symposium on Operating Systems Design and
               Implementation (OSDI 23)},
  pages     = {663--679},
  year      = {2023},
  publisher = {USENIX Association}
}

@inproceedings{agrawal2024sarathi,
  author    = {Amey Agrawal and Nitin Kedia and Ashish Panwar and Jayashree Mohan
               and Nipun Kwatra and Bhargav Gulavani and Alexey Tumanov and
               Ramachandran Ramjee},
  title     = {Taming {Throughput-Latency} Tradeoff in {LLM} Inference with
               {Sarathi-Serve}},
  booktitle = {18th USENIX Symposium on Operating Systems Design and
               Implementation (OSDI 24)},
  pages     = {117--134},
  year      = {2024},
  publisher = {USENIX Association}
}

@inproceedings{sun2024llumnix,
  author    = {Biao Sun and Ziming Huang and Hanyu Zhao and Wencong Xiao and
               Xinyi Zhang and Yong Li and Wei Lin},
  title     = {Llumnix: Dynamic Scheduling for Large Language Model Serving},
  booktitle = {18th USENIX Symposium on Operating Systems Design and
               Implementation (OSDI 24)},
  pages     = {173--191},
  year      = {2024},
  publisher = {USENIX Association}
}

@article{jiang2026turboserve,
  author  = {Youhe Jiang and Haoxu Wang and Haotong Bao and Kai Jiang and
             Jianfei Chen and Jun Zhu and Fangcheng Fu and Jintao Zhang},
  title   = {{TurboServe}: Serving Streaming Video Generation Efficiently
             and Economically},
  journal = {arXiv preprint arXiv:2606.19271},
  year    = {2026}
}

@inproceedings{adya2016slicer,
  author    = {Atul Adya and Daniel Myers and Jon Howell and Jeremy Elson and
               Colin Meek and Vishesh Khemani and Stefan Fulger and Pan Gu and
               Lakshminath Bhuvanagiri and Jason Hunter and Roberto Peon and
               Larry Kai and Alexander Shraer and Arif Merchant and Kfir Lev-Ari},
  title     = {Slicer: {Auto-Sharding} for Datacenter Applications},
  booktitle = {12th USENIX Symposium on Operating Systems Design and
               Implementation (OSDI 16)},
  pages     = {739--753},
  year      = {2016},
  publisher = {USENIX Association}
}

@inproceedings{bykov2011orleans,
  author    = {Sergey Bykov and Alan Geller and Gabriel Kliot and
               James R. Larus and Ravi Pandya and Jorgen Thelin},
  title     = {Orleans: Cloud Computing for Everyone},
  booktitle = {Proceedings of the 2nd ACM Symposium on Cloud Computing
               (SoCC '11)},
  year      = {2011},
  publisher = {ACM}
}

@misc{orleansActivationCount,
  author       = {{Microsoft Orleans}},
  title        = {{ActivationCountBasedPlacement}},
  howpublished = {Microsoft Orleans Documentation},
  year         = {2026},
  note         = {Placement strategy based on recently active grain counts}
}

@inproceedings{adya2010centrifuge,
  author    = {Atul Adya and John Dunagan and Alec Wolman},
  title     = {Centrifuge: Integrated Lease Management and Partitioning
               for Cloud Services},
  booktitle = {7th USENIX Symposium on Networked Systems Design and
               Implementation (NSDI 10)},
  pages     = {1--16},
  year      = {2010},
  publisher = {USENIX Association}
}

@article{mitzenmacher2001power,
  author  = {Michael Mitzenmacher},
  title   = {The Power of Two Choices in Randomized Load Balancing},
  journal = {IEEE Transactions on Parallel and Distributed Systems},
  volume  = {12},
  number  = {10},
  pages   = {1094--1104},
  year    = {2001},
  doi     = {10.1109/71.963420}
}

@inproceedings{ousterhout2013sparrow,
  author    = {Kay Ousterhout and Patrick Wendell and Matei Zaharia and Ion Stoica},
  title     = {Sparrow: Distributed, Low Latency Scheduling},
  booktitle = {Proceedings of the 24th ACM Symposium on Operating Systems
               Principles (SOSP '13)},
  pages     = {69--84},
  year      = {2013},
  publisher = {ACM},
  doi       = {10.1145/2517349.2522716}
}

@inproceedings{schwarzkopf2013omega,
  author    = {Malte Schwarzkopf and Andy Konwinski and Michael Abd-El-Malek
               and John Wilkes},
  title     = {Omega: Flexible, Scalable Schedulers for Large Compute Clusters},
  booktitle = {Proceedings of the 8th ACM European Conference on Computer
               Systems (EuroSys '13)},
  pages     = {351--364},
  year      = {2013},
  publisher = {ACM}
}

@inproceedings{kreps2011kafka,
  author    = {Jay Kreps and Neha Narkhede and Jun Rao},
  title     = {Kafka: A Distributed Messaging System for Log Processing},
  booktitle = {Proceedings of the NetDB Workshop},
  year      = {2011}
}

@inproceedings{kakivaya2018servicefabric,
  author    = {Gopal Kakivaya and Lu Xun and Richard Hasha and
               Shegufta Bakht Ahsan and Todd Pfleiger and Rishi Sinha and
               Anurag Gupta and Mihail Tarta and Mark Fussell and Vipul Modi and
               Mansoor Mohsin and Ray Kong and Anmol Ahuja and Oana Platon and
               Alex Wun and Matthew Snider and Chacko Daniel and Dan Mastrian and
               Yang Li and Aprameya Rao and Vaishnav Kidambi and Randy Wang and
               Abhishek Ram and Sumukh Shivaprakash and Rajeet Nair and
               Alan Warwick and Bharat S. Narasimman and Meng Lin and
               Jeffrey Chen and Abhay Balkrishna Mhatre and Preetha Subbarayalu and
               Mert Coskun and Indranil Gupta},
  title     = {Service Fabric: A Distributed Platform for Building
               Microservices in the Cloud},
  booktitle = {Proceedings of the 13th EuroSys Conference},
  year      = {2018},
  publisher = {ACM},
  doi       = {10.1145/3190508.3190546}
}

@book{lamport2002specifying,
  author    = {Leslie Lamport},
  title     = {Specifying Systems: The {{TLA+}} Language and Tools for
               Hardware and Software Engineers},
  publisher = {Addison-Wesley},
  year      = {2002}
}

@inproceedings{guo2026saga,
  author    = {Dongxin Guo and Jikun Wu and Siu-Ming Yiu},
  title     = {{SAGA}: Workflow-Atomic Scheduling for AI Agent Inference
               on GPU Clusters},
  booktitle = {Proceedings of the 35th International Symposium on
               High-Performance Parallel and Distributed Computing (HPDC '26)},
  year      = {2026},
  publisher = {ACM},
  doi       = {10.1145/3806645.3807598}
}

@inproceedings{mishra2024falcon,
  author    = {Pritish Mishra and Nelson Bore and Brian Ramprasad and
               Myles Thiessen and Moshe Gabel and Alexandre Da Silva Veith and
               Oana Balmau and Eyal de Lara},
  title     = {Falcon: Live Reconfiguration for Stateful Stream Processing
               on the Edge},
  booktitle = {IEEE/ACM Symposium on Edge Computing (SEC 2024)},
  pages     = {234--248},
  year      = {2024},
  publisher = {IEEE},
  doi       = {10.1109/SEC62691.2024.00026}
}

@misc{mishra2024pam,
  author       = {Pritish Mishra and Oana Balmau and Eyal de Lara},
  title        = {{PAM}: Fast Reactive Reconfiguration for Stateful
                  Stream Processing},
  howpublished = {EuroSys 2024 Poster},
  year         = {2024}
}

@inproceedings{qing2025drrs,
  author    = {Yunfan Qing and Wenli Zheng},
  title     = {Towards Fine-Grained Scalability for Stateful Stream
               Processing Systems},
  booktitle = {2025 IEEE 41st International Conference on Data Engineering
               (ICDE)},
  pages     = {3835--3848},
  year      = {2025},
  publisher = {IEEE},
  doi       = {10.1109/ICDE65448.2025.00286}
}

@inproceedings{liu2025syncanimation,
  author    = {Yujian Liu and Shidang Xu and Jing Guo and Dingbin Wang and
               Zairan Wang and Xianfeng Tan and Xiaoli Liu},
  title     = {SyncAnimation: A Real-Time End-to-End Framework for
               Audio-Driven Human Pose and Talking Head Animation},
  booktitle = {Proceedings of the Thirty-Fourth International Joint Conference
               on Artificial Intelligence (IJCAI)},
  pages     = {1657--1665},
  year      = {2025},
  doi       = {10.24963/ijcai.2025/185}
}

@article{liu2025livatar,
  author  = {Haiyang Liu and Xiaolin Hong and Xuancheng Yang and Yudi Ruan and
             Xiang Lian and Michael Lingelbach and Hongwei Yi and Wei Li},
  title   = {{Livatar-1}: Real-Time Talking Heads Generation with
             Tailored Flow Matching},
  journal = {arXiv preprint arXiv:2507.18649},
  year    = {2025}
}

@inproceedings{huang2025liveavatar,
  author    = {Yubo Huang and Hailong Guo and Fangtai Wu and Weiqiang Wang and
               Shifeng Zhang and Shijie Huang and Qijun Gan and Lin Liu and
               Ruihua Huang and Sirui Zhao and Enhong Chen and Jiaming Liu and
               Steven Hoi},
  title     = {Live Avatar: Streaming Real-Time Audio-Driven Avatar
               Generation with Infinite Length},
  booktitle = {European Conference on Computer Vision (ECCV)},
  year      = {2026}
}

@article{li2026hallolive,
  author  = {Chunyu Li and Jiaye Li and Ruiqiao Mei and Haoyuan Xia and
             Hao Zhu and Jingdong Wang and Siyu Zhu},
  title   = {Hallo-Live: Real-Time Streaming Joint Audio-Video Avatar
             Generation with Asynchronous Dual-Stream and Human-Centric
             Preference Distillation},
  journal = {arXiv preprint arXiv:2604.23632},
  year    = {2026}
}

@article{feng2026streamdiffusionv2,
  author  = {Tianrui Feng and Zhi Li and Shuo Yang and Haocheng Xi and
             Muyang Li and Xiuyu Li and Lvmin Zhang and Keting Yang and
             Kelly Peng and Song Han and Maneesh Agrawala and Kurt Keutzer and
             Akio Kodaira and Chenfeng Xu},
  title   = {{StreamDiffusionV2}: A Streaming System for Dynamic and
             Interactive Video Generation},
  journal = {Proceedings of Machine Learning and Systems},
  volume  = {8},
  pages   = {1656--1669},
  year    = {2026}
}
